\PassOptionsToPackage{sort&compress}{natbib}
\documentclass[final,5p]{elsarticlev33}
\usepackage{amsmath,amssymb,amsfonts,amsthm,makeidx,graphicx,booktabs}
\usepackage{cuted}
\usepackage{fancyhdr}
\usepackage{newtxtext,newtxmath}
\usepackage{flushend}
\usepackage{stfloats}
\usepackage{array}
\newcolumntype{L}[1]{>{\raggedright\arraybackslash}p{#1}}
\newcolumntype{R}[1]{>{\raggedright\arraybackslash}p{#1}}

\usepackage{etoolbox}
\AtBeginEnvironment{thebibliography}{\sffamily}

\usepackage{xcolor}
\definecolor{myredbrown}{rgb}{0.55, 0.0, 0.0}
\definecolor{myblue}{rgb}{0.1, 0.2, 0.6}
\usepackage[colorlinks=true]{hyperref}
\AtBeginDocument{%
  \hypersetup{
    linkcolor = myblue, 
    citecolor = myredbrown, 
    urlcolor  = myblue      
  }%
}

\fancypagestyle{elsevierhead}{%
  \fancyhf{}%
  \fancyhead[R]{Collective flavor conversion in dense neutrino plasmas}%
  \fancyhead[L]{D.~F.~G.~Fiorillo}%
  \fancyfoot[C]{\thepage}%
}

\usepackage{microtype}
\usepackage{needspace}

\newcommand{\glossaryheading}[1]{%
  \par\Needspace{6\baselineskip}%
  \medskip
  \textit{#1}\par\nobreak\smallskip
}

\newcommand{\glossarymajorheading}[1]{%
  \par\Needspace{5\baselineskip}%
  \vspace{0.7em}%
  \textbf{#1}\par\nobreak\smallskip
}

\makeatletter
\def\@textbottom{\vskip \z@ \@plus 1pt}
\let\@texttop\relax
\makeatother
\makeatletter
\fancypagestyle{pprintTitle}{%
  \fancyhf{}%
  \fancyhead[R]{Neutrinos from core-collapse supernovae}%
  \fancyhead[L]{Raffelt, Janka, Fiorillo}%
  \fancyfoot[C]{\thepage}%
}
\makeatother

\makeatletter
\providecommand\@combinedblfloats{}
\providecommand\@setmarks{}
\makeatother

\newcommand{\bp}{{\bf p}}
\newcommand{\bv}{{\bf v}}
\newcommand{\br}{{\bf r}}
\newcommand{\bu}{{\bf u}}
\newcommand{\bn}{{\bf n}}
\newcommand{\bK}{{\bf K}}
\newcommand{\bk}{{\bf k}}
\newcommand{\bq}{{\bf q}}

\long\def\exclude#1{}

\usepackage{caption}
\usepackage{tocloft}
\begin{document}

\begin{frontmatter}

\title{\textsf{\textbf{Collective flavor conversion in dense neutrino plasmas}}
}

\tnotetext[t1]{To be published in {\em Encyclopedia of Nuclear Physics} (Elsevier)}

\author[AddressA,AddressB]{Damiano F.~G.~Fiorillo}\ead{damiano.fiorillo@gssi.it}

\address[AddressA]{Gran Sasso Science Institute, Viale F. Crispi 7, L’Aquila, 67100, Italy}
\address[AddressB]{Istituto Nazionale di Fisica Nucleare (INFN), Sezione di Napoli, Complesso Universitario di Monte Sant’Angelo, Via Cintia, 80126
Napoli, Italy}

\begin{abstract}

Compact transient sources, such as supernovae (SNe) and neutron star mergers (NSMs), host a population of thermal neutrinos in their inner cores, which decouple from the dense matter as they stream out. Yet, collisionless does not mean free; their density is large enough to mediate collective waves driven by the coherent weak neutrino--neutrino interaction. These collective waves carry little energy, but crucially they transport flavor. Neutrinos thus form a collisionless plasma with flavor transport driven by flavor waves, whose quanta are called flavomons $\psi$. The wavelength of these flavomons is much shorter than the characteristic scales of SNe and NSMs, so that a brute-force numerical treatment of the neutrino plasma is beyond our capabilities. Yet flavomons cannot be neglected, as the stimulated neutrino decays, schematically $\overline{\nu}_e\to \overline{\nu}_\mu\psi$, cause a rapid buildup in their population, more conventionally called a flavor instability. We review the theory of the neutrino plasma, and systematically discuss the instabilities leading to flavomon growth, and connect this emerging quasiparticle description with recent advances in numerical neutrino flavor kinetics.
\smallskip
\end{abstract}


\end{frontmatter}

\thispagestyle{empty}

\tableofcontents

\newpage

\pagestyle{elsevierhead}

\section{Introduction}\label{sec:introduction}

\noindent Within the Standard Model (SM), neutrinos are the paradigmatic feebly interacting particles, participating only in weak and gravitational interactions, and behaving in most environments independently of each other. Yet the SM predicts neutrino--neutrino interactions. Although no direct observational evidence of their effects exists, at sufficiently high densities these interactions can make neutrinos behave collectively as a plasma, exchanging flavor through the macroscopic weak field that they themselves source, with striking astrophysical consequences. To understand it, we first get a sense of the scales involved.

The environments we have in mind are very dense and hot stellar cores, formed in collapsing massive stars before they explode as supernovae (SNe) or in the remnant of the merger of two neutron stars (NSMs). We focus on the former for definiteness. At densities close to nuclear ones, ($\rho\sim10^{14}-10^{15}\,\mathrm{g\,cm^{-3}}$), and large temperatures ($T\sim30\,\mathrm{MeV}$), neutrinos are trapped and thermalize with the surrounding matter. Electron (anti-)neutrinos equilibrate through charged-current processes with neutrons $n$ and protons $p$ 
\begin{equation}
\nu_e+n\leftrightarrow e^-+p,
\qquad
\bar\nu_e+p\leftrightarrow e^++n.
\end{equation}
Muon neutrinos undergo similar reactions more rarely since muons are scarcer due to their larger masses.
Instead, all flavors interact through neutral-current scattering and pair-production processes, including
\begin{equation}
N+N\leftrightarrow N+N+\nu_\alpha+\bar\nu_\alpha,
\end{equation}
where ($\alpha=e,\mu,\tau$) labels the neutrino flavor. Neutrino-nucleus elastic scattering offers the dominant opacity for heavy-lepton neutrinos $\nu_\mu$ and $\nu_\tau$, although it is inefficient for energy equilibration, and cannot provide chemical equilibration. So, deep in the core of SNe and NSMs, a thermal population of neutrinos of all flavors is established by weak interactions, progressively decoupling from the medium as the density and temperature decrease at radii of order tens of kilometres. The dynamical timescale for free neutrino escape is thus around tens of $\mu$s.

Even at larger radii, where the incoherent interactions with the medium have faded out, neutrinos cannot be treated as noninteracting. As predicted by the SM, they possess mutual interactions, driven by the Hamiltonian
\begin{equation}\label{eq:four_fermion_Hamiltonian}
    \mathcal{H}_{\rm int}=\frac{G_F}{\sqrt{2}}\sum_{\alpha,\beta}\overline{\nu}_{\alpha L} \gamma^\mu \nu_{\alpha L} \overline{\nu}_{\beta L} \gamma_\mu \nu_{\beta L},
\end{equation}
where $G_F$ is the Fermi constant and $\nu_{\alpha L}$ is the left-handed neutrino field. Incoherent neutrino--neutrino collisions have a small rate
\begin{equation}
\Gamma_{\nu\nu}
\sim
G_F^2 n_\nu E_\nu^2
\sim
5\,\mathrm{s^{-1}}
\left(\frac{n_\nu}{10^{32}\,\mathrm{cm^{-3}}}\right)
\left(\frac{E_\nu}{10\,\mathrm{MeV}}\right)^2.
\end{equation}
The corresponding mean free path,
\begin{equation}
\lambda_{\nu\nu}
\sim
6\times10^4\,\mathrm{km}
\left(\frac{10^{32}\,\mathrm{cm^{-3}}}{n_\nu}\right)
\left(\frac{10\,\mathrm{MeV}}{E_\nu}\right)^2,
\end{equation}
is much larger than the region from which neutrinos escape. 

There is however another, much faster timescale associated with neutrino--neutrino interactions. The Hamiltonian in Eq.~\ref{eq:four_fermion_Hamiltonian} shows that each neutrino feels a potential energy sourced by all the others of the order of
\begin{equation}
    \mu=\sqrt{2}G_F n_\nu\simeq 0.6\,\mathrm{cm}^{-1}\,\left(\frac{n_\nu}{10^{32}\,\mathrm{cm}^{-3}}\right).
\end{equation}

This potential energy is negligible in comparison with the neutrino kinetic energy, so it can safely be neglected in the kinetic neutrino motion. For flavor evolution, however, it corresponds to microscopic length and time scales compared with the hydrodynamical evolution.
This concern becomes concrete when coupled with Pantaleone's seminal observation~\cite{Pantaleone:1992eq} that neutrinos which are not in flavor eigenstates, with a nonzero mixed density $\langle \overline{\nu}_{\alpha L} \gamma^\mu \nu_{\beta L}\rangle$, produce, through the Hamiltonian in Eq.~\ref{eq:four_fermion_Hamiltonian}, a potential triggering flavor mixing, often called an off-diagonal refractive index. In practice, a flavor-mixed neutrino background can trigger another one to mix as well.

Even with this insight, rapid flavor dynamics need not trivially ensue. Neutrinos are produced in flavor eigenstates. The mass-induced mixing, measured by the vacuum frequency, is
\begin{equation}
    w_E=\frac{\delta m^2}{2E}\simeq 2\times 10^{-6}\,\mathrm{cm}^{-1}\,\left(\frac{\delta m^2}{10^{-3}\,\mathrm{eV}^2}\right)\,\left(\frac{E_\nu}{10\,\mathrm{MeV}}\right)^{-1},
\end{equation}
where we focus on the heaviest mass splitting $\delta m^2$. This mixing is largely blocked by the larger weak potential energy sourced by electrons with number density $n_e$
\begin{equation}
    \lambda=\sqrt{2}G_F n_e\simeq 60\,\mathrm{cm}^{-1}\,\left(\frac{n_e}{10^{34}\,\mathrm{cm}^{-3}}\right).
\end{equation}
So in principle, there should not be any off-diagonal refractive index produced by these flavor-diagonal neutrinos.

This situation closely resembles a collisionless electronic plasma. Binary collisions may be irrelevant, and an initially neutral plasma produces no macroscopic electric field. Nevertheless, a small perturbation can generate a field that moves additional particles in phase, reinforcing the original perturbation and producing a \textit{plasma instability}. Its nature is most simply described through the elementary, wave-like excitations of the plasma. The quanta of these density waves, mediated by the electric field, are called plasmons $\gamma^*$. A beam--plasma instability, for example, arises from energetic beam particles releasing free energy by stimulated emission of plasmons $e^-\to e^-+\gamma^*$.

A dense neutrino gas can undergo the analogous phenomenon. Pantaleone’s off-diagonal refractive potential is a coherent weak field: it vanishes in a flavor-diagonal state, conserves the total flavor carried by the ensemble, but can redistribute flavor among neutrinos. This field supports collective flavor waves, whose quanta we call \textit{flavomons}~\cite{Fiorillo:2025npi}. A flavor instability may then be viewed as the stimulated emission of flavomons, through which the flavor stored in the angular and energy distributions of the neutrinos is transferred to collective waves. The resulting collective flavor conversion (CFC) may profoundly affect SNe and NSMs, the only astrophysical environments where neutrinos dominate energy and lepton-number transport. In SNe, CFC directly impacts the dynamics of the explosion, by altering the energy deposited by neutrinos, as well as the neutrino signal expected from the next galactic SN; in NSMs, CFC can affect the electron fraction of the ejected material, with vast consequences on the nucleosynthesis of heavy elements.

Dense neutrinos thus create a collisionless plasma mediated by coherent weak fields, much as an electronic plasma is mediated by electromagnetic fields, and a quark-gluon plasma by color fields\footnote{Nowadays quark-gluon plasma broadly denotes deconfined QCD matter, not necessarily in the collisionless state. In its original sense, however, as recently recalled by Shuryak~\cite{Shuryak:2025byj}, the term was ``not a name but a statement'': it expressed the presence of genuine collective plasma phenomena, including screening, plasmon excitations, and Landau damping. For the same reason, \emph{neutrino plasma} should be understood not merely as a name, but as a statement about the collective behavior of the medium.}. The microscopic wavelengths and periods of its collective excitations pose a fundamental obstacle to numerical modeling: simulations cannot simultaneously resolve these scales and the macroscopic evolution of the astrophysical environment. Three questions therefore organize the subject: under what conditions do flavor instabilities arise? How do they saturate? And how can the coupled evolution of neutrinos and flavor waves be described in practice?

Here we review the efforts to answer these questions. In practice, this task is quite complex, as the concept of neutrino plasma is a completely new interpretation of a large body of literature with varying languages and paradigms across decades. We deliberately treat the system as a plasma, which gives a physical account of how flavor instabilities arise and saturate, since it isolates the physical degrees of freedom, namely the waves. Several reviews already cover the more conventional formulation~\cite{Tamborra:2020cul,Volpe:2023met,Johns:2025mlm,Raffelt:2025wty}, which is still discussed here, given the rapid progress in the field, in the attempt of connecting it with the plasma framework.

After introducing the general kinetic equations for neutrino flavor in Sec.~\ref{sec:kinetic_equation}, we discuss in Sec.~\ref{sec:theory_plasma} the general framework that describes this system as a plasma. We later apply this framework, in Sec.~\ref{sec:instabilities} to describe the general instabilities of a collisionless neutrino plasma, and in Sec.~\ref{sec:flavomon_kinetics} to discuss the subsequent evolution of flavomons produced by instabilities. In Sec.~\ref{sec:history}, we present a summary of the progress in the field, attempting to bridge the many viewpoints that have succeeded one another. In Sec.~\ref{sec:numerical}, we review the many fruitful results obtained from numerical solutions of the neutrino kinetic equation; these results are the fundamental empirical evidence that any theory of the neutrino plasma should try to explain. Finally, in Sec.~\ref{sec:outlook} we outline the main open questions for the future.

\section{The kinetic equation for neutrino flavor evolution}\label{sec:kinetic_equation}

Collective phenomena are best described in the kinetic framework. For neutrinos, this means introducing a density matrix $\rho_{\bp,\alpha\beta}(\br)$, describing their flavor state (with $\alpha=e,\mu,\tau$) as a function of position $\br$ and momentum $\bp$. This quasi-classical description is well-justified since the wavelength of neutrinos is of the order of $E_\nu^{-1}\sim 20\,\mathrm{fm}$ for typical neutrino energies of 10~MeV, much shorter than even the centimeter-scale oscillations of flavor waves. Formally, the density matrix is defined in the Wigner sense as
\begin{equation}
    \rho_{\bp,\alpha\beta}(\br)
 =\sum_\bq\,
 e^{i\bq\cdot\br}
 \left\langle
 a^\dagger_{\beta,\bp-\bq/2}
 a_{\alpha,\bp+\bq/2}
 \right\rangle;
\end{equation}
for antineutrinos, we have
\begin{equation}
    \bar\rho_{\bp,\alpha\beta}(\br)
 =\sum_\bq\,
 e^{i\bq\cdot\br}
 \left\langle
 b^\dagger_{\alpha,\bp-\bq/2}
 b_{\beta,\bp+\bq/2}
 \right\rangle,
\end{equation}
where we denote by $a_{\alpha,\bp}$ and $b_{\alpha,\bp}$ the annihilation operators for neutrinos and antineutrinos respectively.

The kinetic equation of neutrinos has been derived along several different lines. The very first derivation in Ref.~\cite{rudzsky1990kinetic} did not include the flavor degree of freedom. The seminal approach of Ref.~\cite{Sigl:1993ctk} is particularly straightforward, starting from the Heisenberg evolution equations of the density operator, but several equivalent and more formal approaches have later been followed, through relativistic Wigner functions~\cite{Sirera:1998ia}, nonequilibrium field theory~\cite{Yamada:2000za,Vlasenko:2013fja,Cirigliano:2014aoa,Kainulainen:2023ocv}, and the BBGKY hierarchy~\cite{Volpe:2013uxl,Serreau:2014cfa,Froustey:2020mcq}. Here we follow the physical derivation in Ref.~\cite{Fiorillo:2024fnl}, which leads to additional terms of physical relevance associated with flavor-evolution-driven neutrino drifts. 

We discuss the whole theory in a two-flavor framework. At a formal level, it is not difficult to generalize it to a three-flavor framework, the only difference being that the SU(2) generators associated with the $2\times 2$ flavor matrices are replaced by the $3\times 3$ generators of SU(3). On the other hand, at the practical level, this makes the formalism more cumbersome and introduces concrete difficulties, such as competing instabilities in different sectors.

Neutrinos in a plasma can be regarded as quasiparticles with a modified energy, due to the weak potential of the medium. This potential is sourced by all particles carrying weak charge. For flavor evolution, however, only its flavor-dependent part is relevant: in ordinary matter this is dominated by charged-current scattering on electrons and muons, although subdominant radiative corrections also contribute. The Lorentz structure of the interaction produces a characteristic angular dependence in the potential between two neutrinos with momenta $\bp$ and $\bp'$ by a factor $1-\bv\cdot \bv'$; here we denote by $\bv=\bp/|\bp|$ the velocity. We also introduce the four-vectors $P^\mu=(E_\bp,\bp)$ with $E_\bp=|\bp|$, and $v^\mu=(1,\bv)$; the latter is denoted as a four-vector purely for convenience, since it does not transform as such.

The effective potential is easily extracted from the interaction Hamiltonian, so we may write the Hamiltonian matrix
\begin{equation}
    {\sf{\Omega}}_\bp=E_\bp+{\sf \Omega}_{\bp,\rm vac}+{\sf\Omega}_{\bp,\rm mat}+{\sf \Omega}_{\bp,\nu\nu};
\end{equation}
here ${\sf\Omega}_{\bp,\rm vac}=\mp w_{E_\bp} (c_V {\sf\sigma_3}-s_V {\sf\sigma_1})/2$ is the vacuum mass splitting (the two signs are for neutrinos and antineutrinos respectively), with $c_V=\cos2\theta_V$ and $s_V=\sin 2\theta_V$ in terms of the mixing angle $\theta_V$, and ${\sf\sigma}_i$ are the Pauli matrices. The matter potential is ${\sf\Omega}_{\bp,\rm mat}=\lambda u\cdot v ({\sf\sigma}_3+1)/2$, where $u^\mu=(1,\bu)$ is the four-component velocity of the electron fluid: this is not a Lorentz four-vector, and becomes the Lorentz four-velocity only in the nonrelativistic limit.  Here we choose a convention such that $\delta m^2$ is always defined positive, so the sign of the ordering is entirely absorbed in the definition of $c_V$; thus $c_V>0$ corresponds to normal mass ordering. Finally, the neutrino--neutrino potential is 
\begin{equation}
    {\sf\Omega}_{\bp,\nu\nu}=\sqrt{2}G_F (\rho^\mu-\overline{\rho}^\mu) v_\mu;
\end{equation}
here we introduce the collective variable $\rho^\mu=\sum_\bp \rho_\bp v^\mu$ and similarly for antineutrinos, where $\sum_\bp=\int d^3\bp/(2\pi)^3$ denotes the usual phase-space sum.
In the following, we will develop the theory as if there were only neutrinos. Antineutrinos are easy to include back; they can be treated as neutrinos with a ``negative'' $w_E$ and with  a negative contribution to $\Omega_{\bp,\nu\nu}$. For this reason, one often introduces the flavor-isospin convention, where antineutrinos are treated with the opposite quantum numbers and $w_E$, so an antineutrino $\overline{\nu}_e$ with a vacuum frequency $w_E$ is identified with a neutrino $\nu_\mu$ with a vacuum frequency $-w_E$.


A particle with Hamiltonian ${\sf\Omega}_\bp$ follows the Hamilton equation, and therefore a collisionless neutrino plasma obeys the same kinetic equation as the excitations of a Fermi liquid~\cite{Landau:1956yop}
\begin{equation}
    \partial_t \rho_\bp+\frac{1}{2}\left\{\partial_\bp{\sf\Omega}_\bp,\partial_\br\rho_\bp\right\}-\frac{1}{2}\left\{\partial_\br {\sf\Omega}_\bp,\partial_\bp\rho_\bp\right\}=i[\rho_\bp,{\sf\Omega}_\bp].
\end{equation}
On the left-hand side, one often keeps only the kinetic term $E_\bp$ in ${\sf\Omega}_\bp$, which is by far the largest one, although as we will see crucial physical effects are missed by this approximation; in fact, even flavor-independent effects, such as gravitational drifts, may be quantitatively relevant. With this choice, the kinetic equation simplifies to its conventional form
\begin{equation}
    (\partial_t+\bv\cdot\partial_\br)\rho_\bp=i[\rho_\bp,{\sf\Omega}_{\bp,\rm vac}+{\sf\Omega}_{\bp,\rm mat}+{\sf\Omega_{\bp,\nu\nu}}].
\end{equation}

This kinetic equation contains in principle the description of the collective dynamics of neutrinos. In spite of its simple form, solving it is a formidable task when flavor instabilities cause the growth of small-scale fluctuations in $\rho^\mu$, as we will see. Since $\rho^\mu$ is determined by the individual density matrices $\rho_\bp$, these equations are intrinsically nonlinear.

We briefly discuss the main assumptions that implicitly underlie this derivation. One of the most controversial is the treatment of the background as a mean field in which individual neutrinos propagate. This amounts to factorizing the many-body Hamiltonian in Eq.~\ref{eq:four_fermion_Hamiltonian} as an effective potential $\langle \overline{\nu}_\alpha \gamma^\mu \nu_\beta\rangle$ self-consistently determined by the evolving neutrino plasma itself. Whether such factorization is reasonable in realistic environments has been hotly debated, as we discuss in more detail in Sec.~\ref{sec:history}. For now, we merely note that quantitatively, for neutrino densities $n_\nu\sim 10^{32}\,\mathrm{cm}^{-3}$, the typical wavelength of excitations described by the kinetic equation is measured by $\mu^{-1}\sim 1\,\mathrm{cm}$. Therefore, within a volume $\mathcal{V}\sim \mu^{-3}$, the number of particles is huge, of the order of $\mathcal{N}\sim 10^{32}$. Representing their collective action by a collective mean field therefore seems sensible, since fluctuations due to individual neutrinos should be suppressed by a factor $\mathcal{N}^{-1/2}$ compared to the mean coherent field.

Beyond the mean-field factorization, there are several other assumptions that implicitly went into the kinetic equation. Firs, there is the question of which mean-field factorization is actually adopted. We have assumed that only the density matrices $\rho_\bp$ and $\overline{\rho}_\bp$ are nonzero, including only coherence between flavors, but not among spins or momenta. In recent years, several works have independently questioned this assumption, proposing that other forms of coherence may spontaneously originate. For example, Sawyer~\cite{Sawyer:2022ugt} has proposed that a neutrino-antineutrino coherence might spontaneously develop $\langle a^\dagger_{\alpha,\bp}b_{\beta,\bp}\rangle$, meaning that some particles turn into a superposition state of a neutrino and an antineutrino. This possibility has been rebutted in Ref.~\cite{Fiorillo:2024wej}, since neutrino--neutrino interactions preserve helicity, while massless neutrinos and antineutrinos have opposite helicities, so such a coherent superposition cannot be spontaneously produced. 

Another possibility is that neutrinos might develop a pairing correlation $\langle a_{\alpha,\bp} b_{\beta,-\bp}\rangle$, as considered originally in Refs.~\cite{Volpe:2013uxl,Serreau:2014cfa}. Such a possibility had been historically discarded since the earliest Ref.~\cite{Sigl:1993ctk}, since the mixed correlator would oscillate very rapidly in time, with a characteristic phase factor $e^{-2iE_\bp t}$, and therefore vanish on average.  It was noticed in Ref.~\cite{Serreau:2014cfa} that the mixed correlator would nonetheless be sourced by the standard neutrino and antineutrino density matrices in an inhomogeneous and anisotropic environment; in other words, it never rigorously vanishes. Nevertheless, the large oscillation frequencies of order $E_\bp$ may still make it extremely small in practice~\cite{Kartavtsev:2015eva}, since the sourcing of the correlator is suppressed by the much slower rate of self-interaction $\mu\ll E_\bp$. Recently, Ref.~\cite{Huang:2026fcb} reconsidered this question, and proposed that this pairing correlator might nevertheless grow unstable: if there are discrete energy levels, neutrinos with the same $E_\bp$ but different directions can still pair, since the rapid phase variation is common to all modes with different directions and can be removed.

The question of pairing correlators becomes much clearer from a physical perspective. A pairing correlator growing unstable means that neutrinos and antineutrinos energetically favor a Bardeen-Cooper-Schrieffer (BCS) state, i.e.~they organize into a correlated state of pairs with opposite momenta~\cite{Fiorillo:2026byi}. Incidentally, it also follows that correlators of the form $\langle a_{\alpha, \bp}a_{\beta,-\bp}\rangle$ and $\langle b_{\alpha,\bp} b_{\beta,-\bp}\rangle$ might equally well develop. The question is whether they are allowed to do so in practice. A pairing instability under a weak attractive interaction can be connected to the general argument put forward by Cooper~\cite{Cooper:1956zz}, as recently noted in Ref.~\cite{Fiorillo:2026byi}. The potential energy among neutrinos is of order $\mu$, while their kinetic energy is of order $E_\bp\gg \mu$. Therefore, a bound state can only form, according to Cooper's criterion, if there is a singular set of states all at the same energy, such as the Fermi surface of electrons in a metal. In this case, all the states lying at the singular surface can pair efficiently because they all have the same kinetic energy, so it takes an infinitesimally small potential energy to restructure neutrinos along the surface. Since neutrinos in hot, dense environments are generally not strongly degenerate---except in the innermost regions where they are in thermal and chemical equilibrium---such conditions are unlikely to arise.

Finally, in the kinetic equations, we neglected the force exerted by the gradient in the potential $\partial_\br {\sf\Omega}_{\bp,\nu\nu}$. As we will see, when a flavor instability occurs, this potential spontaneously becomes inhomogeneous over length scales comparable with $\mu^{-1}$, so that we expect a true force acting on neutrinos of the order $F\sim \mu^2$. Over timescales of order $\tau\sim\mu^{-1}$, this alters the momentum of neutrinos by an amount of order $F\tau \sim \mu$. In other words, in an unstable neutrino plasma, neutrinos are accelerated and decelerated by flavor conversions, an effect first noted in Ref.~\cite{Fiorillo:2024fnl}. From the perspective of the actual neutrino kinetic energies, this is a small effect, since $\mu \ll E_\bp$. However, it has the very practical consequence that the potential interaction energy, i.e.\ $\mathcal{H}_{\rm int}$, is \textit{not} conserved in an inhomogeneous plasma, since it can be exchanged with the much larger reservoir of kinetic energy. This includes in particular any form of unstable plasma; instabilities always involve small-scale perturbations, which therefore mediate energy exchange. Thus, neglecting the force terms introduced in Ref.~\cite{Fiorillo:2024fnl} leads to equations that effectively do not conserve energy, by an amount that is not concerning for hydrodynamical evolution, but precludes a consistent tracking of potential interaction energy.

\section{Dielectric response of a neutrino plasma}\label{sec:theory_plasma}

\subsection{Kinetic variables and field variables}

We now turn to the field itself $\sqrt{2} G_F\rho^\mu$ in terms of the total current, describing the weak potential sourced by the neutrino plasma. We distinguish this rather clearly from the kinetic variables $\rho_\bp$, since the latter depend on $\bp$, the momenta of individual neutrino modes. Instead, the field variables are integrated over the entire population, so they are coarse-grained information.

This potential is completely analogous to the electric potential in a plasma. The ``Poisson equation'', which links the electric potential to the charge in the plasma, takes here a trivial form: $\sqrt{2} G_F \rho^\mu$ is simply proportional to the total charge density $\rho^\mu=\sum_\bp\rho_\bp v^\mu$, which is why we denote the field directly by $\rho^\mu$. This simplicity is a direct consequence of the short-range nature of weak interactions: the potential is, up to a factor, the local ``charge'' density. The resulting neutrino-plasma theory, in the form presented here, has been developed in a series of works~\cite{Fiorillo:2024bzm,Fiorillo:2024uki,Fiorillo:2024pns,Fiorillo:2025ank,Fiorillo:2025npi,Fiorillo:2025zio,Fiorillo:2025kko}.

We usually are interested in the response to a small off-diagonal flavor perturbation, so we consider a system in which neutrinos are nearly all in a flavor-diagonal state, with a small fraction of neutrinos in a superposition state. We do this by explicitly separating the density matrix $\rho_\bp$ as
\begin{equation}
    \rho_\bp=\frac{1}{2}\begin{pmatrix}
        n_\bp+D_\bp && \psi^*_\bp\\
        \psi_\bp && n_\bp-D_\bp
    \end{pmatrix}=\frac{n_\bp}{2}+\frac{\vec{P}_\bp\cdot \vec{\sigma}}{2},
\end{equation}
where we have introduced the total occupation number $n_\bp$ and the so-called polarization vector $\vec{P}_\bp$\footnote{We denote by arrows the vectors in flavor-isospin space, while vectors in real space are denoted by bold-face characters.}; its $z$ component $D_\bp$ is the difference between electron and muon neutrino occupation number, while its off-diagonal component $\psi_\bp=P^x_\bp+iP^y_\bp$ encodes the degree of coherence between flavors. We use a similar expansion for the total field 
\begin{equation}
    \rho^\mu=\frac{1}{2}\begin{pmatrix}
        n^\mu+D^\mu && \psi^{*,\mu}\\
        \psi^\mu && n^\mu-D^\mu
    \end{pmatrix}.
\end{equation}

\subsection{Theory of classical flavor waves}\label{sec:classical_theory_flavor_waves}

First we consider a purely classical treatment of the flavor field. Our perturbative assumption amounts to considering $\psi_\bp$ and $\psi^\mu$ as small in comparison with the on-diagonal parts. One could of course at a formal level expand around any equilibrium state, not necessarily flavor-diagonal~\cite{Johns:2025yxa}. At a practical level, this has however not been done, both because it is much harder to do, and because a clear case for a macroscopic nondiagonal flavor configuration in astrophysical contexts has not yet emerged.

For $\psi_\bp$, the neutrino kinetic equations become
\begin{equation}\label{eq:eqs_motion}
    iv\cdot\partial\psi_\bp=(w_E c_V-\lambda u\cdot v-\sqrt{2}G_F D\cdot v)\psi_\bp+\sqrt{2}G_F \psi\cdot v D_\bp+w_E s_V D_\bp.
\end{equation}
The parenthesis acts as an effective frequency for each of the individual neutrino modes $\bp$, so in some sense they are the ``energies'' of the noninteracting neutrino modes; they include the matter potential sourced by electrons and neutrinos, as well as the vacuum frequency. The second term $\sqrt{2} G_F \psi\cdot v D_\bp$ is the most interesting one, corresponding to an off-diagonal field $\psi$ triggering flavor mixing in the individual mode $\bp$. Finally, the last term $w_E s_V D_\bp$ describes the unavoidable off-diagonal source due to the nonzero mixing angle, which remains nonzero even if all $\psi_\bp=0$. This term plays the role of a seeding, acting as a small external forcing of off-diagonal coherence. Notice that it is the vacuum mixing angle entering the kinetic equations. 

An interesting feature is that the mixing term $w_E s_V D_\bp$ appears as a source term, similar to an additional external field $\psi_{\rm ext}^\mu$, albeit with an energy-dependent coefficient. Thus, for generality, let us rewrite the equations formally as if a total field $\tilde{\psi}^\mu=\psi^\mu + \psi^\mu_{\rm ext}$ is slowly applied
\begin{equation}
    iv\cdot\partial \psi_\bp=(w_E c_V-\lambda u\cdot v-\sqrt{2}G_F D\cdot v)\psi_\bp+\sqrt{2}G_F \tilde{\psi}\cdot v D_\bp.
\end{equation}
How do neutrinos in the plasma respond to a slowly inserted field? Since the response to the field is linear, we may consider a single monochromatic wave $\tilde{\psi}^\mu\propto \mathrm{lim}_{\eta\to 0^+}\, e^{-i\Omega t+i\bK\cdot \br+\eta t}=e^{-iK\cdot X+0 t}$; the infinitesimal $\eta$ is introduced to describe the slow insertion of the field. For conciseness, we simply denote it by a $+0$. Then the response of the plasma is
\begin{equation}
    \psi_\bp=\frac{\sqrt{2}G_F \tilde{\psi}\cdot v D_\bp}{(K+\sqrt{2}G_F D+\lambda u)\cdot v - w_E c_V +i 0}.
\end{equation}
It is conventional to define a shifted $k^\mu=K^\mu+\Lambda^\mu$, where $\Lambda^\mu=\sqrt{2}G_F D^\mu+\lambda u^\mu$ is the total potential sourced by both electrons and neutrinos.
In turn, the responding neutrinos will produce a new field of their own
\begin{equation}
     \psi^\mu=\sum_\bp \psi_\bp v^\mu=\chi^{\mu}_\nu \tilde{\psi}^\nu.
\end{equation}
While the mathematical description may seem involved, physically the mechanism is identical to standard electrodynamics: the external field $\tilde{\psi}$ evokes a response from the particles, which produce another field, analogous to the polarization of a dielectric, mediated by the ``flavor susceptibility''
\begin{equation}
    \chi^{\mu}_\nu=\sum_\bp \frac{\sqrt{2}G_F D_\bp v^\mu v_\nu}{k\cdot v-w_E c_V+i0}.
\end{equation}
Purely for notational convenience, we note that several works consider axially symmetric angular distributions and focus on modes directed along the axis of symmetry $\bn$; in this case, the integral over the azimuthal angle around this axis factors out, and so it is quite useful to find in the literature a distribution written in terms of vacuum frequency $w_E$ and polar angle $v=\bv\cdot\bn$ as
\begin{equation}
    D_{v,w_E}=\frac{E_\bp^2}{4\pi^2}\frac{dE_\bp}{dw_E}D_\bp=\frac{\delta m^6}{32\pi^2w_E^4}D_\bp,
\end{equation}
defined so that $\sum_\bp D_\bp=\sum_{v,w_E} D_{v,w_E}=\int dv dw_E D_{v,w_E}$. We will not focus on these special cases here.

We can now return to the physical condition that the external field should be ultimately just the field produced by the neutrinos themselves, combined with the small mixing field $ \psi^\mu_{\rm ext}$
\begin{equation}
    \psi^\mu=\chi^\mu_\nu(\psi^\nu+\psi^\nu_{\rm ext}).
\end{equation}

The inhomogeneous term $\chi^\mu_\nu \psi^\nu_{\rm ext}$ can in fact be expressed through the seeding term in Eq.~\ref{eq:eqs_motion}: a simple calculation shows that, in terms of the integrated seeding
\begin{equation}\label{eq:seeding}
    \mathcal{S}^\mu=\sum_\bp \frac{w_E s_V v^\mu }{k\cdot v - w_E c_V+i0} \int d^4 X e^{iK\cdot X} D_\bp.
\end{equation}
Here the integral over $X^\mu$ is performed to extract the Fourier component at $K^\mu$ from the background distribution $D_\bp$. If $D_\bp$ is truly homogeneous and static, then only the homogeneous mode $\bK=0$ can be sourced; in practice, some small-scale fluctuations are likely present, so that there will be power even at $|\bK|\sim \mu \epsilon$. Eq.~\ref{eq:seeding} is also non-local in time, i.e.\ it depends on the behavior of $D_\bp$ at all times, since it provides the amplitude for a mode with a given frequency $\Omega_\bK$. In the End Matter of Ref.~\cite{Fiorillo:2026tee}, the same result is obtained by a more physical, local treatment in time using an initial-value approach.

After defining the ``flavor dielectric tensor'' $\varepsilon_{\mu\nu}=\delta_{\mu\nu}-\chi_{\mu\nu}$, we have
\begin{equation}\label{eq:dielectric_function_definition}
    \varepsilon^\mu_\nu \psi^\nu=\mathcal{S}^\mu.
\end{equation}
Eq.~\ref{eq:dielectric_function_definition} is the central element of the neutrino plasma theory. 

In the same vein as for standard electrodynamics, we see that something special happens when $\varepsilon^\mu_\nu(K)$ has a zero, i.e.\ a vector such that $\varepsilon^\mu_\nu(K) e^\nu=0$. Indeed, in this case, there are special solutions that sustain themselves without the need for an external action. These are called ``flavor waves'', exactly like plasma waves in a medium. Most importantly, if a solution is found such that $\Omega=\Omega_R+i\Omega_I$ has a positive imaginary part, it corresponds to the existence of waves that not only sustain themselves, but that grow exponentially in time $e^{-i\Omega t}\propto e^{\Omega_I t}$. In this way, we connect this general description with the conventional linear stability analysis~\cite{Izaguirre:2016gsx} testing for unstable solutions.

An interesting point is connected with the magnitude of the seeding for flavor waves. Their frequencies are determined by the dispersion relation $\varepsilon_{\mu\nu}(K) e^\nu=0$. The typical scale of their temporal frequencies will therefore be $\omega_\bk\sim \mu \epsilon$, which is the largest dimensional parameter. On the other hand, the physical frequency $\Omega_\bK\sim \lambda$ if $\lambda \gg \mu \epsilon$, as is the case below the shock wave of SNe. Since $\lambda$ is a huge frequency in comparison with the hydrodynamical rate of change of the background, in the integral in Eq.~\ref{eq:seeding} we need to extract the Fourier power of a slowly varying function at a large frequency $\Omega_\bK$. If neutrinos instantaneously fill the plasma volume, then the resulting integral from Eq.~\ref{eq:seeding} is of the type
\begin{equation}
    \int_0^{+\infty} dt' e^{i\Omega_\bK t'}D_\bp = \frac{D_\bp}{i(\Omega_\bK+i\epsilon)}. 
\end{equation}

Since $\Omega_\bK\sim \lambda$, it follows that the amplitude of the flavor wave is proportional to $s_V/\lambda$. This conclusion is the root of the historical argument that a large matter term results in a suppressed mixing angle. At the theoretical level, its appearance was justified only in the case of the homogeneous instability connected with the bipolar pendulum~\cite{Hannestad:2006nj}, and discussed in Sec.~\ref{sec:history}. Our argument here, first presented in a similar, although less general, form in Ref.~\cite{Fiorillo:2024pns}, shows how this approximations emerges beyond the single homogeneous mode. Many works today adopt therefore the approximation of neglecting the matter term altogether, and using in its place a suppressed mixing angle; an empirical study of this approximation is presented in Ref.~\cite{Shalgar:2025oht}. 

However, our theoretical argument sheds new light on the question, showing several shortcomings: first, the appearance of the factor $\Omega_\bK^{-1}$ is entirely connected with the history of the flavor evolution, since it depends on the assumption of $D_\bp$ becoming non-zero at $t=0$. Numerical simulations usually make this assumption, but realistic evolution need not follow it. Second, $\lambda$ only affects the normalization of the flavor wave, which depends on $\sin\theta_V$; however, the dispersion relation itself also depends on $w_E \cos\theta_V$, which is \textit{not} affected by $\lambda$. Therefore, while changing the vacuum mixing angle to an effective one may capture the reduction in the flavor wave amplitude, it also alters the frequency splitting appearing in the dispersion relation, leading to the wrong growth rate. Finally, when matter density is inhomogeneous, the effect of matter goes far beyond that of reducing the flavor wave amplitude, as we discuss in more detail in Sec.~\ref{sec:flavomon_propagation}. A more complete discussion of the matter term and its relation to the effective mixing angle is given in Ref.~\cite{Fiorillo:2024pns}.

\subsection{Quantum theory---flavomons}

\begin{figure}
    \includegraphics[width=\columnwidth]{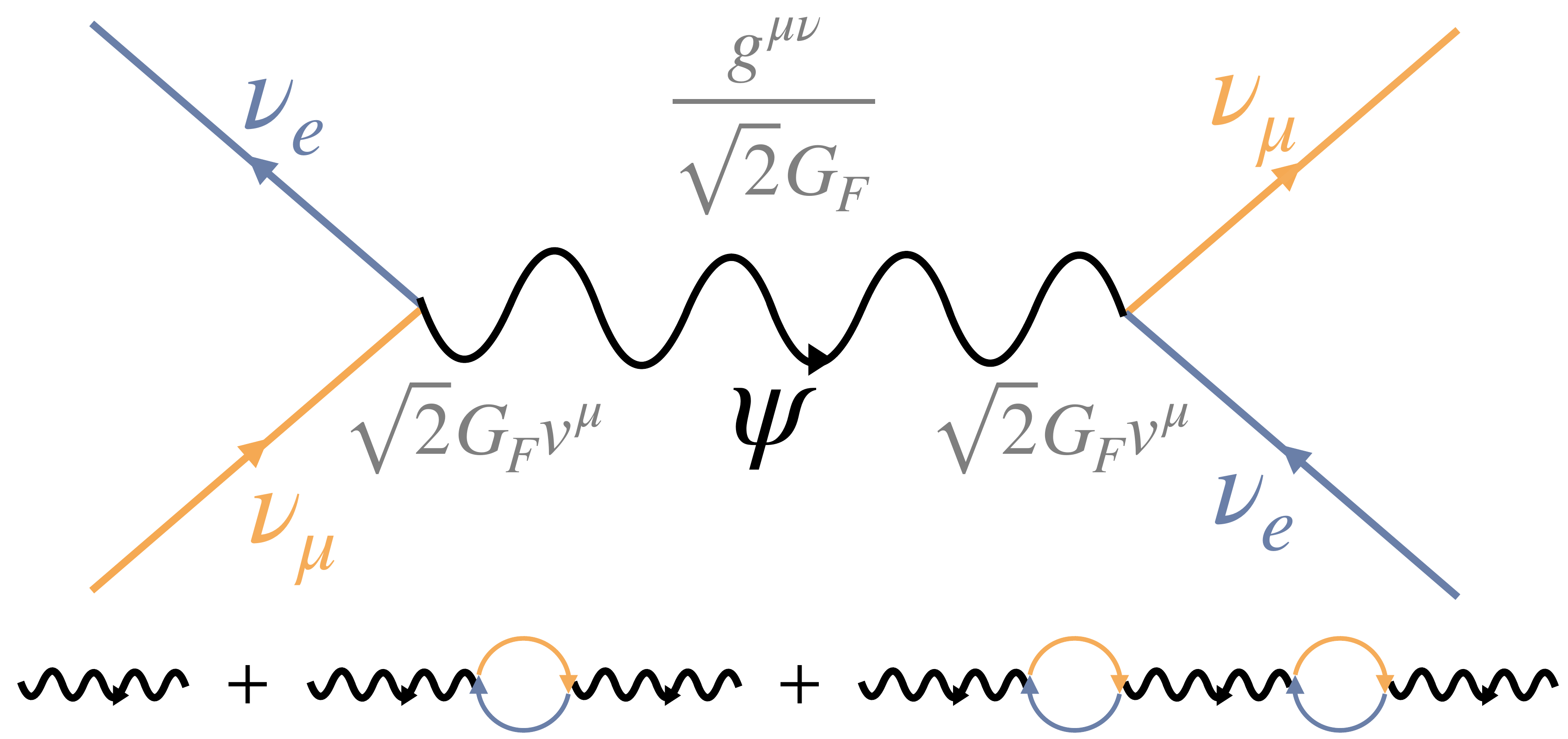}
    \caption{Diagrammatic representation of neutrino--neutrino contact interaction, and its medium-renormalized, dynamical interaction mediated by flavomons. At tree level the contact interaction has no dynamics, so its amplitude is independent of exchanged energy and momentum. The random-phase-approximation, or bubble resummation, shown in the figure endows it with dynamics, encoded in the flavor dielectric function.}\label{fig:flavomons}
\end{figure}

\begin{figure}
    \includegraphics[width=\columnwidth]{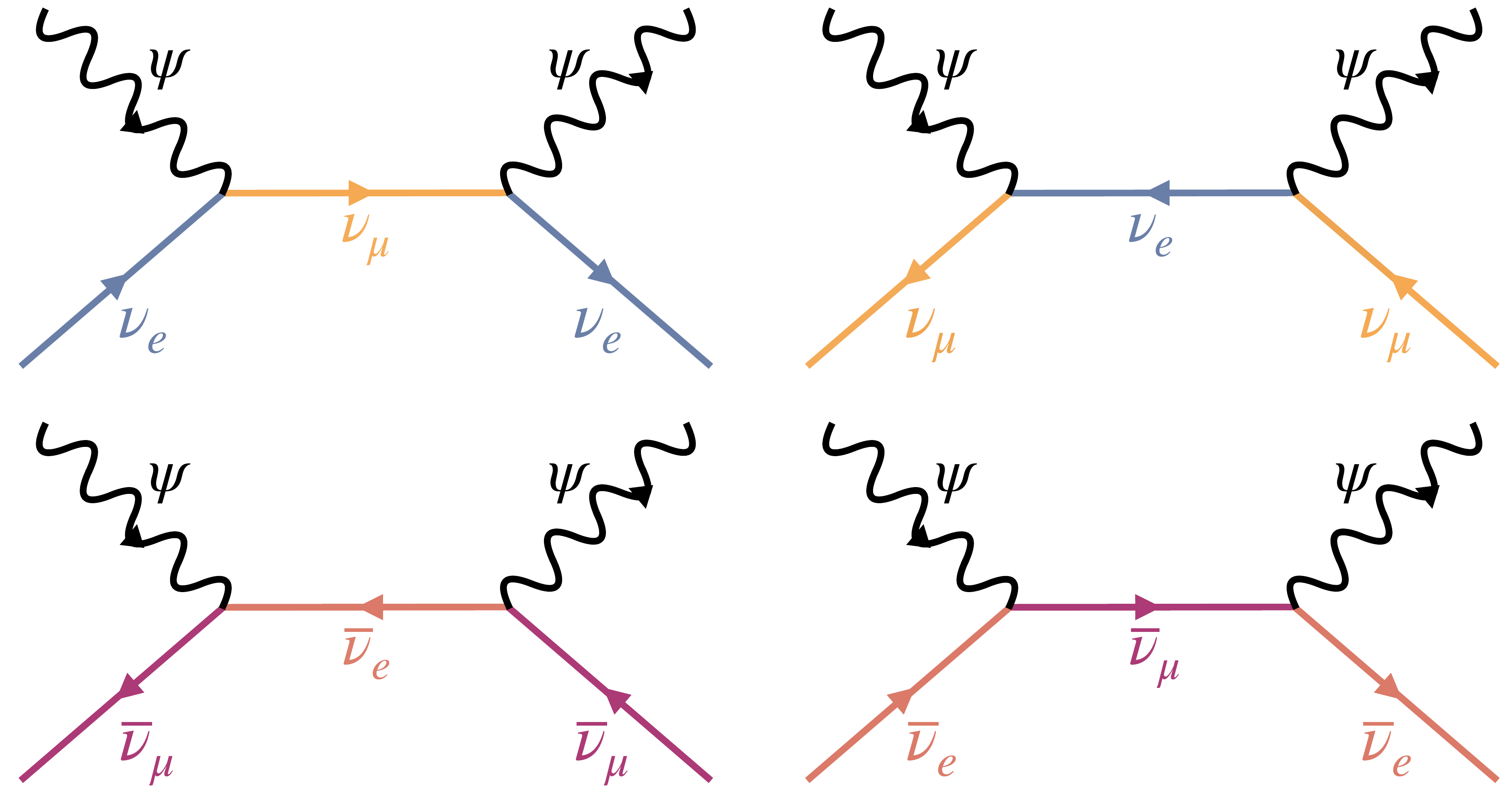}
    \caption{Flavomon refraction due to forward scattering from neutrinos of different flavors. In the main text, to simplify notation, we only include neutrinos without antineutrinos. The arrows denote the direction of the lepton number, so for antineutrinos they run in the opposite direction to the particle, as conventional in relativistic field theory.}\label{fig:flavomon_refraction}
\end{figure}

The classical description provides a simple picture: flavor waves are self-sustaining oscillations of the flavor field. There is however another viewpoint that sheds more light on their nature. The Hamiltonian in Eq.~\ref{eq:four_fermion_Hamiltonian} describes an interaction among two neutrinos. At the diagrammatic level, we may represent it as a flavor exchange (Fig.~\ref{fig:flavomons}).

While in vacuum neutrino--neutrino interaction is static and local, in the medium flavor exchange acquires dynamics. Just as for any other boson, we may define a self-energy induced by the interaction with a neutrino bubble diagram. Physically, these ``flavomons'', which in vacuum are purely static degrees of freedom, in the medium acquire dynamics, because they can be coherently absorbed and re-emitted by the neutrinos in the medium. Therefore, while in vacuum flavor exchange is associated with a static propagator $D_{\mu\nu}=g_{\mu\nu}/\sqrt{2} G_F$, in the medium this becomes a new, emergent species with a propagator $\mathcal{D}^{\nu\lambda}$
\begin{equation}
    (D_{\mu\nu}^{-1}-\Sigma_{\mu\nu})\mathcal{D}^{\nu\lambda}=\delta^\lambda_\mu.
\end{equation}
The self-energy associated with the coherent interaction with neutrinos is exactly equal to the classical flavor susceptibility $\Sigma_{\mu\nu}=\sqrt{2}G_F\chi_{\mu\nu}$.  Therefore, in terms of the flavor dielectric function, the flavomon propagator is
\begin{equation}
    \mathcal{D}_{\mu\nu}=\frac{\varepsilon^{-1}_{\mu\nu}}{\sqrt{2}G_F}.
\end{equation}
Therefore, a zero of the dielectric tensor corresponds to a pole of the propagator; as signaled by the classical theory, this zero implies  a propagating degree of freedom, i.e.\ a quasiparticle. In the vicinity of this pole $\Omega\simeq \Omega_\bK$, we may expand the dielectric tensor projected on the physical eigenvector $e^\nu$, so
\begin{equation}
    \mathcal{D}_{\mu\nu}\simeq \frac{e_\mu e_\nu \mathcal{Z}}{\sqrt{2}G_F(\Omega-\Omega_\bK)},
\end{equation}
where $\mathcal{Z}=(e^\alpha e^\beta \partial_\Omega \varepsilon_{\alpha\beta})^{-1}$ is the wavefunction renormalization. 

Compared to the standard quantum field theory in vacuum, in a medium we have the subtler feature that the dispersion relation $\Omega_\bK$ is determined by the medium itself. Since the neutrino medium is out-of-equilibrium, there is no universal dispersion relation independent of the background distribution, as it exists instead e.g.\ for plasmons in an equilibrium electronic plasma. A particularly troublesome point is that the frequency $\Omega_\bK$ and the wavefunction renormalization $\mathcal{Z}$ may both be either positive or negative. Within the usual conventions of quantum field theory, positive (negative) energies are associated with flavomons (antiflavomons) for positive $\mathcal{Z}$. For negative $\mathcal{Z}$, the situation is specular.

Thus flavomons are the quanta of the classical flavor field $\psi^\mu$ that was introduced in Sec.~\ref{sec:classical_theory_flavor_waves}. 
To match with their definition as $\psi^\lambda=2\rho_{\mu e}^\lambda=2\sum_\bp \langle a^\dagger_{e,\bp}a_{\mu,\bp}\rangle v^\lambda$, a flavomon must be defined such that it carries the quantum number of a $\nu_\mu \overline{\nu}_e$ pair. In the literature this quantum number is often called the difference in lepton number (DLN); in a three-flavor plasma there would be three different kinds of DLN ($e\overline{\mu}$, $\mu\overline{\tau}$, $e\overline{\tau}$). The notion of DLN is usually introduced as a convenient mathematical tool to describe the difference between the $\nu_e$ and the $\nu_\mu$ distributions which appear in the dispersion relation. (In fact, since most studies are done in two flavors, this quantity has also been variously called electron lepton number (ELN) or electron minus $x$-neutrino lepton number (E-XLN).) The flavomon paradigm shows clearly that the DLN has instead a physical meaning: it is the charge of the flavomon.
The field admits of course also antiflavomons with the opposite quantum numbers. Therefore, a flavomon can be emitted by $\nu_\mu\to \nu_e \psi$ or $\overline{\nu}_e\to \overline{\nu}_\mu\psi$; an antiflavomon can be emitted by $\nu_e\to \nu_\mu\overline{\psi}$ or $\overline{\nu}_\mu\to \overline{\nu}_e \overline{\psi}$.

Since flavomons carry flavor, or rather DLN, we may compare them directly with gluons, which mediate color exchange among quarks through their color charge. The analogy with gluons also makes the extension to three flavors immediately clear. With three flavors, rather than having a single flavomon species carrying lepton number $\nu_\mu \overline{\nu}_e$, we would have three flavomon species (plus their antiflavomons). These correspond to the six color-changing gluon states. The additional flavor-diagonal states are called neutrino-plasmons in Ref.~\cite{Fiorillo:2025npi}; they do not participate in the linear growth of flavor instabilities, hence we do not address them here, but they are essential for the nonlinear evolution, since they can be produced by flavomon--flavomon interactions.

An even more illuminating comparison is with magnons, the quanta of spin waves in magnets. Classically, a spin wave is a collective precession of spins propagating through matter like a wave. At the quantum level, such a wave is an ensemble of magnons which carry spin 1; the emission of a spin wave then corresponds to the spin flip of a single electron. The analogy with flavomons is thus complete, with the two ``isospin'' states of neutrinos identified with their $\nu_e$ and $\nu_\mu$ states. Just like an individual magnon, created in a sea of ``spin up'', can be regarded as an itinerant ``spin down'', so a flavomon, created in a sea of $\nu_e$, can be regarded as an itinerant $\nu_\mu$.

Beyond the powerful visual analogy, the flavomon concept allows one to immediately understand the origin of the flavor susceptibility. As we have discussed, the self-energy of flavomons comes from their coherent forward scattering on neutrinos, shown by the two processes in Fig.~\ref{fig:flavomon_refraction}. Let us compute the rate for this process explicitly, using the Feynman rules illustrated in Fig.~\ref{fig:flavomons}. The unperturbed energy of electron and muon neutrinos, as we have seen, is
\begin{equation}
    \varepsilon_\bp=E_\bp+s\frac{\Lambda\cdot v-w_E c_V}{2}=E_\bp+s\frac{\Delta E_\bp}{2},
\end{equation}
where the sign $s=+1$ for $\nu_e$ and $\overline{\nu}_\mu$, and $s=-1$ for $\nu_\mu$ and $\overline{\nu}_e$; we neglect flavor-independent and momentum-independent energies which are irrelevant. Therefore, for the first diagram, if the initial electron neutrino has momentum $\bp$ and energy $\varepsilon_\bp=E_\bp+\Delta E_\bp/2$, the intermediate neutrino has momentum $\bp-\bK$ and energy $\varepsilon_{\bp-\bK}=E_{\bp-\bK}-\Delta E_{\bp-\bK}/2\simeq E_\bp-\bK\cdot\bv-\Delta E_\bp/2$. Therefore, the intermediate neutrino has a ``virtuality'', i.e.\ it is off-shell by the energy
\begin{equation}\label{eq:virtuality}
    \Delta \varepsilon=\varepsilon_{\bp-\bK}+\Omega_{\bK}-\varepsilon_\bp\simeq \Omega_{\bK}-\bv\cdot\bK+\Delta E_\bp=k\cdot v-w_E c_V.
\end{equation}

Thus we can easily compute the forward-scattering amplitude, which we identify as minus the self-energy as usual, by standard perturbation theory. This immediately gives
\begin{equation}
    -i\Sigma^{\mu\nu}=-i\sum_\bp\frac{2G_F^2(n_{\nu_e,\bp}-n_{\nu_\mu,\bp})v^\mu v^\nu}{k\cdot v-w_E c_V+i0},
\end{equation}
confirming as we expected that $\Sigma^{\mu\nu}=\sqrt{2}G_F \chi^{\mu\nu}$. We see that in this language the presence of the DLN in the numerator has a physical meaning that was completely opaque in the classical approach; it is the difference between the amplitude for flavomon forward-scattering from $\nu_\mu$ and $\nu_e$. Antineutrinos, according to the flavor-isospin convention, naturally contribute to these refractive processes.

Most importantly, the characteristic denominator in the flavor susceptibility also acquires a new meaning; it is the propagator of the intermediate neutrino. If this denominator can vanish, it implies that a neutrino can emit or absorb a flavomon on-shell. We now turn to this possibility.

\subsection{Landau damping of flavomons}

\begin{figure}[t!]
    \includegraphics[width=\columnwidth]{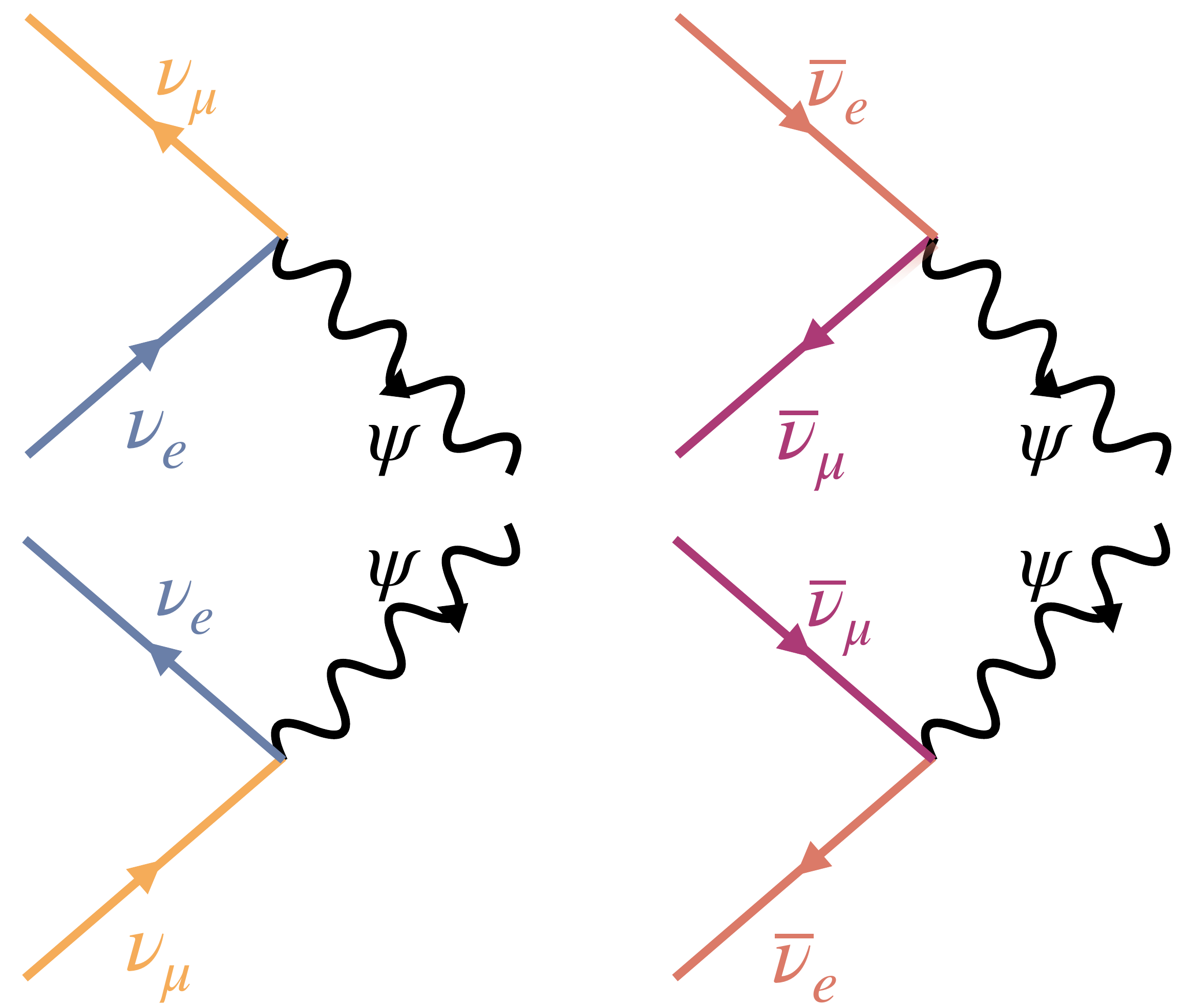}
    \caption{Flavomon absorption and emission processes. In the main text, to simplify notation, we only include neutrinos without antineutrinos.}\label{fig:flavomon_emission}
\end{figure}

If the denominator of the flavor susceptibility can vanish for real frequencies, this implies dissipation. In this case, dissipation means that a flavomon can be absorbed or emitted on-shell. The vanishing denominator causes no divergence in the susceptibility, since we have the well-defined $i0$ prescription (first introduced in this context in Refs.~\cite{Fiorillo:2023mze,Fiorillo:2023hlk,Fiorillo:2024bzm}). This retarded prescription comes from causality---the flavomon field is inserted adiabatically, so the response can only follow, not precede, this insertion.

The rate of absorption of a flavomon is easily obtained from our Feynman rules; a flavomon is absorbed by $\nu_e$ and emitted by $\nu_\mu$. The relevant processes are collected in Fig.~\ref{fig:flavomon_emission}. Thus, a flavomon mode with wavevector $K^\mu$, with an occupation number $N_\bK$, evolves according to a kinetic equation

\begin{multline}
\partial_t N_{\bK}
=
\sum_{\bp} w(\bK,\bp)
\Bigl[
n_{\nu_\mu,\bp+\bK}
(1-n_{\nu_e,\bp})(1+N_{\bK})
\\
{}-
n_{\nu_e,\bp}N_{\bK}
(1-n_{\nu_\mu,\bp+\bK})
\Bigr].
\label{eq:flavomon_emission}
\end{multline}

where
\begin{equation}
    w(\bK,\bp)=\sqrt{2}G_F |\mathcal{Z}| 2\pi \delta[k\cdot v-w_E c_V]|e\cdot v|^2
\end{equation}
is the transition kernel of a flavomon with momentum $\bK$ from an electron neutrino with momentum $\bp$, obtained from the Feynman rules; here the zero component of $k^\mu$ should be interpreted only as the real part $\omega_{\bk,R}$ rather than the full complex frequency. Critically, flavomon emission from $\nu_\mu$ is stimulated by the Bose factor $1+N_\bK$.

Since $|\bK|\sim \mu\ll |\bp|$, we can safely neglect $\bK$ in comparison with $\bp$. Furthermore, in the limit of large flavomon occupation number $N_\bK\gg 1$,  we have $\partial_t N_\bK=2\gamma_\bK N_\bK$, where 

\begin{multline}
\gamma_{\bK}
=
\frac{G_F|\mathcal Z|}{\sqrt{2}}
\sum_{\bp}
\bigl(n_{\nu_\mu,\bp}-n_{\nu_e,\bp}\bigr)
\\
{}\times
2\pi\delta\bigl(k\cdot v-w_Ec_V\bigr)
|e\cdot v|^2 .
\label{eq:emission_rate}
\end{multline}
The factor of $2$ in the definition of $\gamma_\bK$ is conventional; if the flavomon field amplitude grows with a rate $\gamma_\bK$, as typically assumed, the flavomon occupation number, proportional to the squared amplitude, grows with twice the rate.

Among all neutrinos, only some are able to satisfy the energy conservation condition. We call these \textit{resonant neutrinos}. If, among these neutrinos, there is a larger amount with positive electron lepton number (i.e.\ $\nu_e$ or, for antineutrinos, $\overline{\nu}_\mu$) compared to the opposite ones, then flavomons are absorbed ($\gamma_\bK<0$). The resulting phenomenon is well-known in plasma physics as Landau damping~\cite{Landau:1946jzi}, and was introduced in the flavor context in Ref.~\cite{Fiorillo:2024bzm}. 

Since $N_\bK$ was assumed to be very large, the quantum nature of flavomons is not essential, yet it strongly guides intuition since the absorption from individual neutrinos flips their flavor, which can only be described as a quantum process. Nevertheless, mathematically the results obtained for $N_\bK\gg 1$ must also follow without reference to the notion of flavomons; indeed, Ref.~\cite{Fiorillo:2024bzm} obtained the same result for the damping rate through an analytical expansion of the dispersion relation.

At the classical level, Landau damping is connected with the resonant absorption of flavor waves by neutrinos. This phenomenon corresponds to the well-known Cherenkov process, whereby a particle moving faster than the phase velocity of the wave can absorb or emit it. Similarly, neutrinos moving faster than the phase velocity of flavomons can absorb them or emit them, because they can move in phase with the wave. This is the reason for the original nomenclature of resonant neutrinos. 

The rate of Landau damping has been derived here under the assumption that flavomons are well-defined excitations, i.e.\ $|\gamma_\bK|\ll |\Omega|$; in fact, formally the requirement is more stringent $|\gamma_\bK|\ll |\Omega\pm K|$. If the damping rate becomes comparable to the eigenfrequency, the flavomon excitation is so short-lived that treating it as a particle is meaningless. In that case, Landau damping survives, but can only be found as a solution of the exact dispersion relation $\varepsilon^\mu_\nu \psi^\nu=0$ with $\Omega_I<0$. In the limit of $|\Omega_I|$ small, this general solution matches Eq.~\ref{eq:emission_rate}. In the latter, we have accounted for a generically continuous energy and angular distribution. On the other hand, observing Landau damping in a concrete setup (e.g.\ by initializing a stable neutrino plasma with a certain configuration for the field $\psi^\nu$ and observing its time evolution) is generally not easy. This is due to the angular integration in $\varepsilon^\mu_\nu$ being performed over a compact range; this finite range causes the appearance of power-law tails at late times that easily mask the exponential suppression in the flavor field due to Landau damping. We do not touch on this topic; the generic time evolution of an initial condition is discussed in Ref.~\cite{Fiorillo:2024bzm}.

\section{Collisionless instabilities as stimulated flavomon emission}\label{sec:instabilities}

\subsection{General mechanism of instability}

When the growth rate turns positive, Landau damping turns to an instability. The nature of this instability can now be understood as the Cherenkov emission of flavor waves, or, what is the same, the decay of neutrinos from one flavor to another with emission of flavomons. We now turn to a statistical understanding of the conditions under which Cherenkov emission may dominate over absorption, leading to an instability.

The general principle is that, for any given flavomon mode, defined by $\bK$ and $\Omega_\bK$, there will only be a special group of neutrinos that are kinematically able to emit or absorb the flavomon, satisfying the resonant condition
\begin{equation}
    k\cdot v=w_E c_V.
\end{equation}
This picture is particularly applicable for weakly interacting flavomons, which behave as excitations with well-defined energies. The more strongly the flavomon interacts, the larger will be the imaginary part of its energy, which allows for considerable deviations from the exact energy conservation condition. However, weakly interacting excitations are physically motivated; in fact, excitations with a large negative $\gamma_\bK$ are strongly damped and therefore unlikely to have any impact on realistic situations. Excitations with a large positive $\gamma_\bK$ are strongly unstable; while they can of course exist, in realistic scenarios we expect instabilities to be driven slowly by some exterior alteration of the neutrino distributions~\cite{Johns:2024dbe,Fiorillo:2024qbl}, and therefore at least at their earliest appearance unstable modes should be weakly interacting with a small growth rate. 

Motivated by these physical considerations, we aim to consider neutrino distributions which are close to stability. To do this, we need to understand when we expect stability in the first place. 

\subsection{Stability of neutrino plasma}\label{sec:stable_plasma}

\begin{figure*}[t!]
    \centering
    \includegraphics[width=\textwidth]{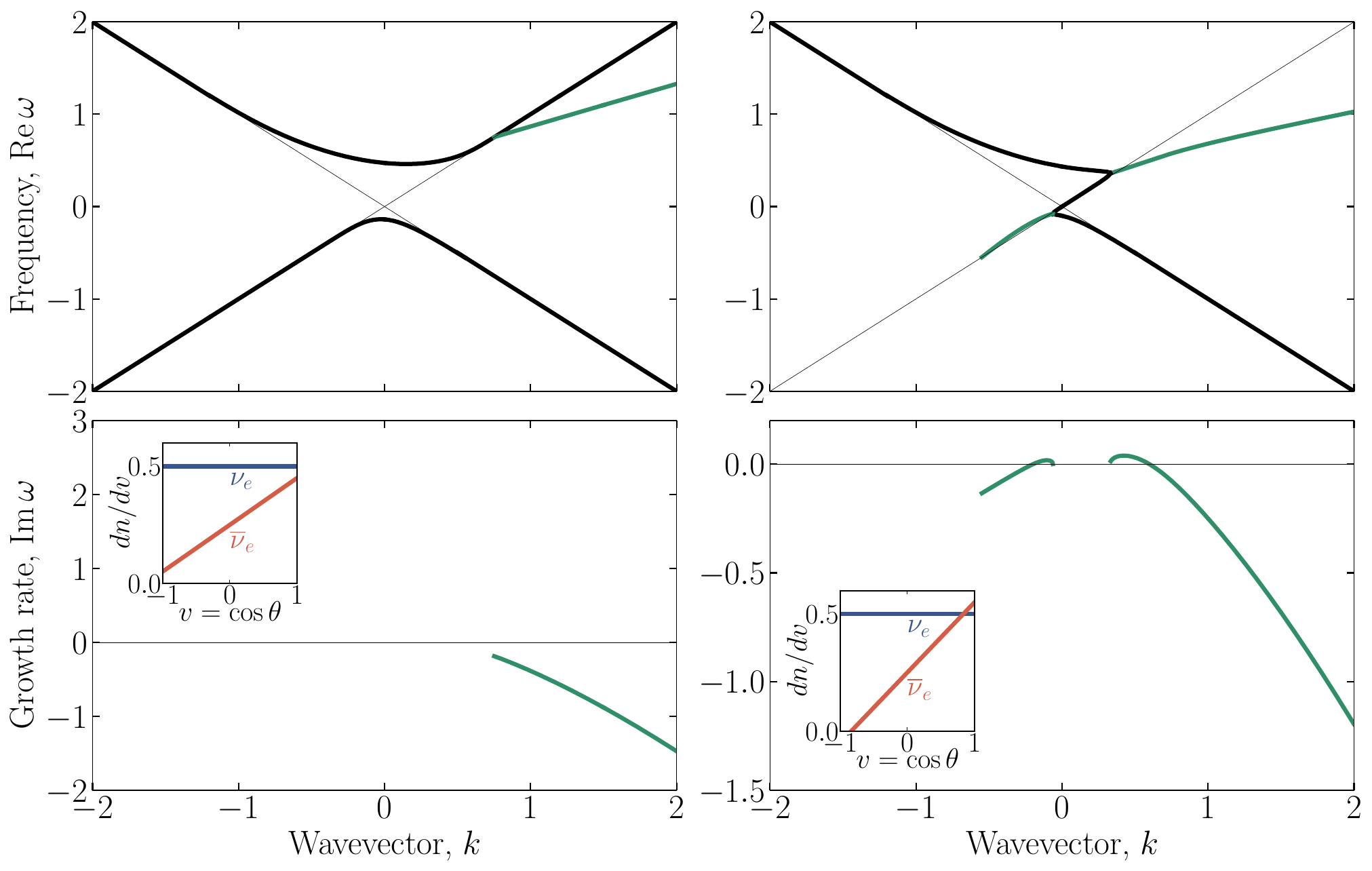}
    \caption{Dispersion curves of longitudinal flavomons moving along the axis of symmetry, for a stable case without angular crossing (left) and an unstable case with angular crossing (right). The angular distributions of $\nu_e$ and $\overline{\nu}_e$ are shown in the inset. Branches with a nonzero growth rate (positive or negative) are shown in green.}
    \label{fig:dispersion_stable}
\end{figure*}

As we have seen, a specific flavomon mode is stable if the resonant neutrinos preferentially absorb rather than emit it. On the other hand, we have provided no argument for which flavor of neutrinos will absorb or emit a flavomon, so that the kinematic condition per se does not help us to determine whether a given neutrino distribution is stable or not. It turns out that, at least for the more limited scope of a sufficient condition for stability, we may rely on a global argument based on the conservation of total lepton number.

The argument itself is indeed quite simple. An instability corresponds to the emission of flavomons, e.g.\ $\overline{\nu}_e\to \overline{\nu}_\mu\psi$. Suppose this is possible, so that an individual $\overline{\nu}_e$ flips its flavor. The total flavor of the plasma must be conserved, which at the level of the microscopic reaction is achieved since $\psi$ carries DLN. In order for this to be possible, the remaining plasma, without the emitting $\overline{\nu}_e$, must be able to support an additional unit of $\mu\overline{e}$ DLN. This, however, is only possible if the remaining plasma contains at least some $\nu_e$ or $\overline{\nu}_\mu$. Indeed, if the rest of the plasma contains only $\overline{\nu}_e$ and $\nu_\mu$, a new unit of DLN cannot be added to it without violating the conservation of flavor. 

This argument can be reformulated in an even more streamlined version: since flavor conversions ultimately are collisionless exchanges of flavor among neutrino modes induced by weak interactions, at least some neutrino modes must have differing flavors among each other, otherwise they cannot possibly exchange flavor. 
Thus, a neutrino plasma is certainly stable if all neutrino modes carry exactly the same DLN, which is usually described as having no crossing. A crossing in this case is defined as a transition, in the energy and angular distribution of neutrinos, from one DLN to another. Antineutrinos in this case can be treated, as we have already discussed, as neutrinos with a negative vacuum frequency $w_E$; so, e.g., a plasma containing only $\nu_e$ and $\overline{\nu}_e$ possesses a crossing, since the $\overline{\nu}_e$ have negative DLN at negative vacuum frequency, while a plasma containing only $\nu_e$ and $\overline{\nu}_\mu$ has no crossing, and is thus stable.
The necessity of a crossing for an instability can actually be proved formally from the mathematical dispersion relation~\cite{Morinaga:2021vmc,Dasgupta:2021gfs}; the argument above was given first for the regime of massless neutrinos in Ref.~\cite{Johns:2024esf}, where it was formulated in the language of flavor isospins---a system with no crossing is in a state of maximum isospin, and thus cannot evolve into any other state without violating isospin---and later generalized for any form of instability in Ref.~\cite{Fiorillo:2024bzm}. 

Thus a neutrino plasma containing only one sign of the DLN for all neutrino and antineutrino modes is stable. An additional stability property can be inferred in the massless neutrino limit, when $w_E\to 0$. In this case, the resonant condition for neutrinos becomes energy-independent $k\cdot v=0$,
and also does not depend on whether the emitting particle is a neutrino or an antineutrino. Therefore, a given flavomon mode is emitted by all neutrinos and antineutrinos moving along a given direction, integrated over energy; in other words, only the energy-integrated DLN, summed over neutrinos and antineutrinos, enters the dispersion relation. It follows immediately that, for an instability to appear, one should have an \textit{angular} crossing, i.e.\ a crossing in the energy-integrated angular distribution of the DLN.

We now have a convenient starting point to examine the potential instabilities of a neutrino plasma. A plasma containing only $\nu_e$ is stable, by the above arguments. In such a system, the possible flavomon states only carry positive DLN, such that they can only be absorbed by $\nu_e$ but not emitted; this ensures the stability of the plasma. If we now add a small amount of neutrinos with the opposite DLN, e.g.\ a small amount of $\overline{\nu}_e$, which we refer to as flipped neutrinos, these will be prone to flavomon emission. We focus first on the massless limit, and consider a case without an angular crossing, where there cannot be any instability.

The typical flavomon dispersion relation is shown in Fig.~\ref{fig:dispersion_stable}, left panel. We focus here only on an axially symmetric neutrino plasma, and consider $\bK$ directed along the axis of symmetry. Thus we have $\bK=K_z \hat{\mathbf{z}}$ and $\bk=k_z\hat{\mathbf{z}}$, so that $\bk\cdot \bv=k_z v$ with $v=\cos\theta$. Their polarization vector $e^\mu$ is also chosen such that its spatial part is along the axis of symmetry; these modes are generally called longitudinal. For simplicity, we set $\lambda=0$; a constant $\lambda$ corresponds simply to a shift in the real part of the flavomon frequency.

The dispersion relation of flavomons in the massless limit consists of two stable branches for which $|\omega_\bk|>|\bk|$; we say that they are above, or outside of, the light cone. Their phase velocity is superluminal, so no neutrino can satisfy the condition for Cherenkov emission. These flavomons are therefore protected from emission or absorption, hence their stability. 

In addition to these stable modes, we also find a branch of modes \textit{below} the light cone. Since their phase velocity is subluminal, these can be kinematically emitted and absorbed; specifically, a flavomon with phase velocity $u=\omega_\bk/k_z$ can be emitted by all neutrinos with $v>u$ for $k_z>0$. Since the angular distribution possesses no crossing, the integrated DLN for $v>u$ is certainly positive---in our example $\nu_e$ dominate at every angle---and so these modes cannot be unstable, in agreement with our general understanding. Therefore, these modes are Landau-damped, as the dispersion relation directly reveals. 

This simple system is already, so to speak, on the brink of instability. It can be made unstable in two different ways. First, a small deformation producing an angular crossing removes the protection from instability even if neutrino masses are neglected. Since the evolution in this limit is often termed fast evolution, we refer to \textit{fast instabilities}. Second, even without an angular crossing, when neutrino mass splittings are included, protection from stability is again removed, and an instability may ensue, although not guaranteed; we refer to these as \textit{neutrino-mass-induced}, or \textit{slow instabilities}.

\subsection{Fast instabilities}\label{sec:fast_instabilities}


We first continue to focus on the fast regime of massless neutrinos. The naming fast is a direct reference to the absence of small parameters in the equations of motion; since $w_E=0$, the relevant timescales are determined purely by the neutrino--neutrino interaction strength $\mu$. We thus consider a slight deviation in the antineutrino angular distribution, such that $\overline{\nu}_e$ dominate over $\nu_e$ near the forward direction. The resulting angular crossing implies that the system is no longer protected from instability. In fact, a more stringent result can be proved, that an angular crossing is not only necessary, but also sufficient, for the appearance of a fast instability. This result was first proven by Morinaga~\cite{Morinaga:2021vmc} based on the mathematical structure of the dispersion relation; we will present later a physical proof developed in Ref.~\cite{Fiorillo:2024bzm}.

The dispersion relation for the modified neutrino distribution which now includes an angular crossing is shown in Fig.~\ref{fig:dispersion_stable}, right panel. The key difference is that the previously Landau-damped branch now evolves into an unstable branch for a limited range of wavenumbers close to the positive light cone $\omega_\bk=k_z$. It is not hard to interpret this behavior physically; a flavomon close to the positive light cone with phase velocity $u=\omega_\bk/k_z<1$ can be emitted or absorbed by all neutrinos with $\cos\theta>u$. As we already discussed, in a $\nu_e$-dominated plasma, flavomon states carry negative DLN, i.e.\ they behave as a particle--hole pair $\overline{\nu}_e\nu_\mu$, so they are emitted by $\overline{\nu}_e$ and absorbed by $\nu_e$. Due to the angular crossing, close to $v\simeq 1$, $\overline{\nu}_e$ dominate over $\nu_e$; therefore, flavomons with $u\simeq 1$ will be preferentially emitted rather than absorbed.
This argument also gives a rather direct prediction, namely that flavomons moving exactly with $\omega_\bk/k_z=v_{\rm cr}$, where $v_{\rm cr}$ is the crossing direction, will lie at the transition between instability and Landau damping, so they will have $\gamma_\bK=0$. This prediction is directly validated by the dispersion relation.

The physical picture we have developed also allows us to give a general proof of the sufficiency of the angular crossing; for this, we do not need to assume axial symmetry. If an angular crossing exists, such that along a certain direction $\bn$ there is vanishing DLN, we can construct a flavomon state moving along that direction $\bk=k\bn$ and having exactly $\omega_\bk=|\bk|$; there will certainly be some value of $|\bk|$ for which this satisfies the dispersion relation. A formal proof of this statement is given in Ref.~\cite{Fiorillo:2024bzm}. Because the flavomon moves with the speed of light, only neutrinos moving exactly in the same direction can emit or absorb it. But since the DLN in this direction vanishes, on balance there will be no emission or absorption, and this mode has zero growth rate $\gamma_\bK=0$. By continuity, if we now consider a flavomon moving towards a slightly different direction, it will be damped on one side of the crossing and grow on the opposite side, showing that an instability must exist. Thus, sufficiency is defined by the specific kinematic structure of flavomon emission and absorption for massless neutrinos, namely by the observation that, for $\gamma_\bK\to 0$, flavomons can only be emitted or absorbed by neutrinos moving faster than its phase velocity. 

For weak instabilities, defined by a vastly dominant DLN (e.g.\ $\nu_e$ in our reference case) and a small number of flipped neutrinos with the opposite DLN (e.g.\ $\overline{\nu}_e$ in our reference case), we can go beyond this argument and even conclude which flavomon directions will correspond to growth, namely those directions resonant with flipped neutrinos, which emit flavomons, rather than absorbing them.

As stressed above, fast evolution of a neutrino plasma is ruled by a single parameter, the neutrino--neutrino interaction strength $\mu$; more precisely, since only the energy-integrated DLN, defined as the difference $n_{\nu_e}-n_{\nu_\mu}-n_{\overline{\nu}_e}+n_{\overline{\nu}_\mu}$ enters, the growth rate $\gamma_\bK$ and flavomon frequency $\omega_\bk$ is determined by the parameter $\mu \epsilon$, where we use

\begin{equation}
    \epsilon\sim \frac{n_{\nu_e}-n_{\nu_\mu}-n_{\overline{\nu}_e}+n_{\overline{\nu}_\mu}}{n_\nu}
\end{equation}
as a schematic reminder that only the DLN determines the fast evolution. The symbol $\sim$ is here used since $\epsilon$ in reality changes along different directions, e.g.\ $\epsilon$ vanishes at an angular crossing. What we imply here is therefore the typical order of magnitude across the entire distribution, that determines the typical timescale of fast evolution. So $\omega_\bk\sim \mu \epsilon$. On the other hand, for shallow angular crossing, where the number of flipped neutrinos in the crossing region is much smaller than the total, the growth rate may be significantly smaller. From the general formula for growth rate of weak instabilities Eq.~\ref{eq:emission_rate}, we see that its typical order of magnitude will be
\begin{equation}
    \gamma_\bK\sim \mu \frac{n_{\nu,\rm flip}}{n_\nu},
\end{equation}
where $n_{\nu,\rm flip}$ is the typical number density of neutrinos involved in the angular crossing.

As the growth rate becomes larger, the flavomon frequency gets broadened ever more significantly. The delta function enforcing conservation of energy in the flavomon emission process in Eq.~\ref{eq:emission_rate} therefore is replaced by a Lorentzian with a width $\gamma_\bK$. In practice, this means that not only neutrinos moving faster than the phase speed of the flavomon---the Cherenkov condition---but all neutrinos can in fact participate in the interaction. The growth rate will therefore depend on \textit{global}, integrated properties, rather than on the \textit{local} ones. A heuristic way to understand the typical growth rate in this regime is to notice that it is not simply proportional to the number of flipped neutrinos, but also to the effective spectral power in the Lorentzian; since its width is $\gamma_\bK$, when evaluated near the center we have
\begin{equation}
    \delta[\Delta \varepsilon]\to \frac{\gamma_\bK/\pi}{\Delta \varepsilon^2+\gamma_\bK^2}\simeq \frac{1}{\pi \gamma_\bK}.
\end{equation}
Therefore, by self-consistency, the growth rate will scale as
\begin{equation}
    \gamma_\bK\sim \mu \frac{n_{\nu,\mathrm{flip}}}{n_\nu}\frac{\mu \epsilon}{\gamma_\bK},
\end{equation}
giving finally
\begin{equation}
    \gamma_\bK\sim \mu \sqrt{\frac{\epsilon n_{\nu,\rm flip}}{n_\nu}}.
\end{equation}
This square-root behavior can be easily confirmed by modeling e.g.\ the angular distribution as a discrete set of beams. Instabilities in these nonresonant regimes are often termed in plasma physics hydrodynamical or \textit{reactive instabilities}.

\subsection{Neutrino-mass-induced (slow) instabilities}

\begin{figure}[t!]
    \centering
    \includegraphics[width=\columnwidth]{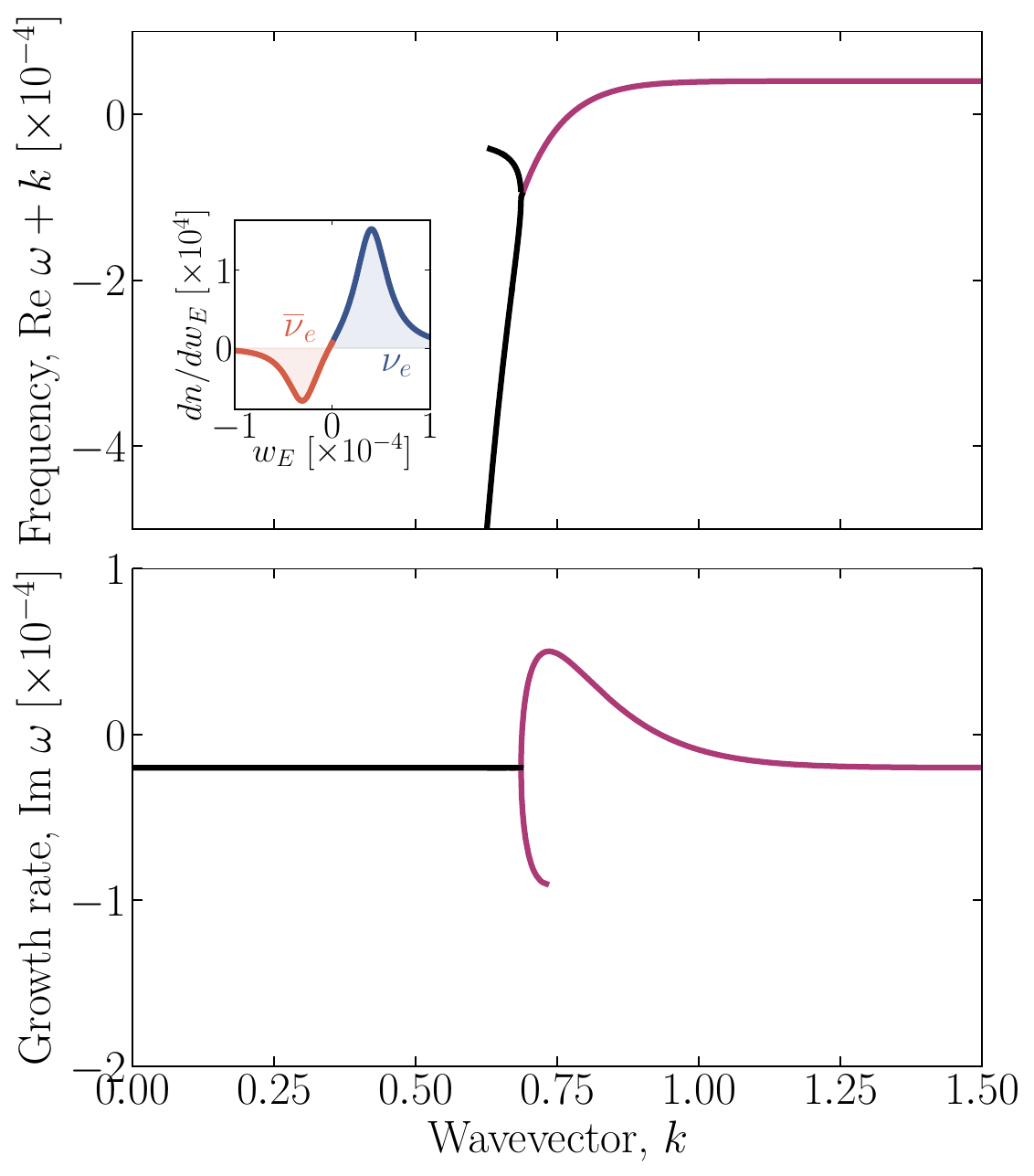}
    \caption{Dispersion curves of the lower longitudinal flavomon branch for the energy distributions, obtained as a superposition of two Lorentzians in frequency $w_E$, shown in the inset. We show the deviation \(\operatorname{Re}\omega+k\) from the negative light cone and the corresponding growth rate. As the mode approaches the light cone within a distance of order \(w_E\), it can be emitted by the flipped antineutrino population and becomes unstable; the interacting branch is shown in magenta. Notice that damped modes can disappear abruptly when they touch one of the singularities of the dispersion relation, explaining the abrupt break in the black and purple curve for $\mathrm{Im}\;\omega<0$. This phenomenon might be absent if the dispersion relation itself has no singularities, as discussed in Ref.~\cite{Kost:2026jrk}.}
    \label{fig:dispersion_slow_unstable}
\end{figure}

We now return to the angular distribution of Fig.~\ref{fig:dispersion_stable}, without an angular crossing. When a nonzero vacuum frequency $w_E$ is included, the system is no longer protected against instability, since neutrinos with different vacuum frequencies now emit or absorb with different rates. Thus, the question follows whether including $w_E$ does indeed induce an instability, by turning some of the existing flavomon branches from stable to unstable.

The Landau-damped flavomons in Fig.~\ref{fig:dispersion_stable} are already strongly damped due to the interaction with the energy-integrated DLN which tends to absorb them. Let us consider instead the stable flavomons outside the light cone, which were previously protected from interacting with neutrinos since the on-shell condition $\omega_\bk-\bv\cdot \bk=0$ could not be fulfilled. With a nonzero vacuum frequency, the kinematic condition for flavomon emission and absorption becomes
\begin{equation}
    \omega_\bk-\bv\cdot\bk-w_E c_V=0.
\end{equation}
The vacuum frequency is a small parameter $w_E$; within the cores of SNe, typically $w_E/\mu\sim 10^{-5}$. For flavomons very close to the light cone, this condition can then be fulfilled. Specifically, if we consider neutrinos nearly collinear with a given flavomon direction at a small angle $\theta$, such that $\bv\cdot \bk=|\bk| \cos\theta\simeq |\bk|(1-\theta^2/2)$, and $\omega_\bk\simeq |\bk|+\chi_R$, the condition becomes
\begin{equation}\label{eq:energy_conservation_light}
    \frac{|\bk|\theta^2}{2}\simeq w_E c_V-\chi_R.
\end{equation}
Thus for a given flavomon mode lying above the light cone by an amount $\chi_R$, all neutrinos with $w_E c_V>\chi_R$ are kinematically able to participate in emission/absorption processes, while those with $w_E c_V<\chi_R$ cannot. \textit{If} the former are dominated by the flipped species, then flavomons will be emitted rather than absorbed, and an instability will ensue. Analogously, we could look for the states of negative energy in Fig.~\ref{fig:dispersion_stable}, for which $\omega_\bk\simeq -|\bk|+\chi_R$; we see easily that these can be emitted or absorbed by neutrinos satisfying $w_E c_V<\chi_R$.

Let us apply this intuition to our reference case, in which we have a dominant population of $\nu_e$ and a subdominant population of $\overline{\nu}_e$. Therefore, the flipped DLN lies at $w_E<0$. For normal ordering ($c_V>0$), we then find that flavomons with negative $\omega_\bk<0$ can be emitted on-shell by antineutrinos primarily, since their kinematic condition is $w_E<\chi_R/c_V$; in particular, a flavomon with $\chi_R<0$ can be emitted \textit{only} by antineutrinos, and thus can only be unstable.

A direct solution of the dispersion relation completely confirms this conclusion, as we show in Fig.~\ref{fig:dispersion_slow_unstable}. We focus only on the lower branch with $\mathrm{Re}(\omega_\bk)<0$, and we show the deviation of the frequency of the light cone. As the flavomons approach the light cone by an amount comparable with the vacuum frequency, the flipped antineutrinos with $w_E<0$ become able to emit them. The growth rate correspondingly is of the order of $\gamma_\bK\sim w_E$, the frequency range of the emitting antineutrinos.

This picture holds in the limit in which flavomons are well-defined excitations; this regime was termed \textit{narrow resonance} in Ref.~\cite{Fiorillo:2025zio}. As we increase the number of flipped neutrinos, the emission rate may become so large that the width $\gamma_\bK\gg w_E$. In this case, the flavomon energy is not well-defined and becomes broadened, so that the kinematic condition need not be fulfilled exactly, but only within an uncertainty of the order of the growth rate. From the energy conservation condition we can already gather that there are two different regimes:

\begin{itemize}
    \item if $w_E\ll \gamma_\bK\ll \mu \epsilon$, the frequency broadening is large enough to allow neutrinos of all frequencies to participate in the interaction. However, it is not large enough to allow neutrinos of all \textit{directions} to do so. The energy cost associated with a neutrino which is not collinear with the flavomon is $k(\cos\theta-1)\sim \mu \epsilon$ for large $\theta$. Therefore, the only neutrinos that can emit a flavomon in this regime are those collinear with the flavomon itself; specifically, from Eq.~\ref{eq:energy_conservation_light} we deduce that only an opening angle $\theta\sim \sqrt{\gamma_\bK/\mu\epsilon}$ is involved. This so-called \textit{broad resonance} regime was analyzed in Refs.~\cite{Fiorillo:2024pns,Fiorillo:2025ank,Fiorillo:2025zio}, with the conclusion that the growth rate scales as $\gamma_\bK\sim w_E/\epsilon$, where $\epsilon$ is the local frequency-integrated DLN along the flavomon direction. The applicability of this regime requires $\gamma_\bK\ll\mu \epsilon$, i.e.\ $w_E\ll \mu \epsilon^2$, so it is only valid for not too small DLN;
    \item as $\epsilon$ decreases further, the total number of flipped neutrinos may become so similar to the number of unflipped ones that $\gamma_\bK\gtrsim \mu \epsilon$. In this case, the flavomon width becomes so large that even neutrinos noncollinear with the flavomon can participate in the interaction. This is the \textit{nonresonant} regime in Ref.~\cite{Fiorillo:2024pns}, for which the growth rate depends on the entire angular distribution rather than only the portion directed towards the flavomon mode, and it scales as $\gamma_\bK\sim \sqrt{w_E \mu}$. We stress that this regime only applies when $\gamma_\bK\gg \mu \epsilon$, so these nonresonant ``slow'' instabilities---in the sense of being induced by the neutrino masses--- in fact typically grow faster than the fast instabilities we have examined earlier. This is in fact a somehow confusing point, in that the scaling $\gamma_\bK\sim \sqrt{w_E \mu}$ has long been associated with slower growth than the fast instabilities because of the small factor $w_E\ll \mu$. However, this regime is reached only when the total DLN $\epsilon$ is so small that fast instabilities are much slower, with a rate determined by $\mu\epsilon$, not $\mu$.
\end{itemize}

Overall, slow instabilities, may play a prominent role in realistic environments simply because of their ubiquity. Fast instabilities may grow faster, but they only appear connected to specific portions of the neutrino angular distribution which exhibit an angular crossing. Slow instabilities, on the other hand, appear with much less stringent requirements.

Note that a one-to-one connection between spectral crossings and instability does not exist~\cite{Fiorillo:2025kko}, differently from fast instabilities which are one-to-one connected with angular crossings. The difference depends on the kinematics of the process. For fast instabilities, a flavomon moving with the speed of light can be emitted by all neutrinos exactly collinear with it; if those neutrinos are predominantly flipped, an instability follows, making the connection transparent. Instead, for slow instabilities, a flavomon moving nearly with the speed of light can be emitted by all neutrinos with frequencies below the threshold: Eq.~\ref{eq:energy_conservation_light} shows that neutrinos with frequencies below the threshold can still emit flavomons at a slightly tilted angle $\theta$. Hence, only the integrated DLN below the kinematic threshold determines whether an instability appears; so a distribution with positive (unflipped) DLN both for $w_E\to \pm\infty$ may not have an instability even if there is a region of negative DLN in between. Such examples are constructed and analyzed in more detail in Ref.~\cite{Fiorillo:2025kko}.

\subsection{Collisional instabilities}

\begin{figure}[t!]
    \includegraphics[width=\columnwidth]{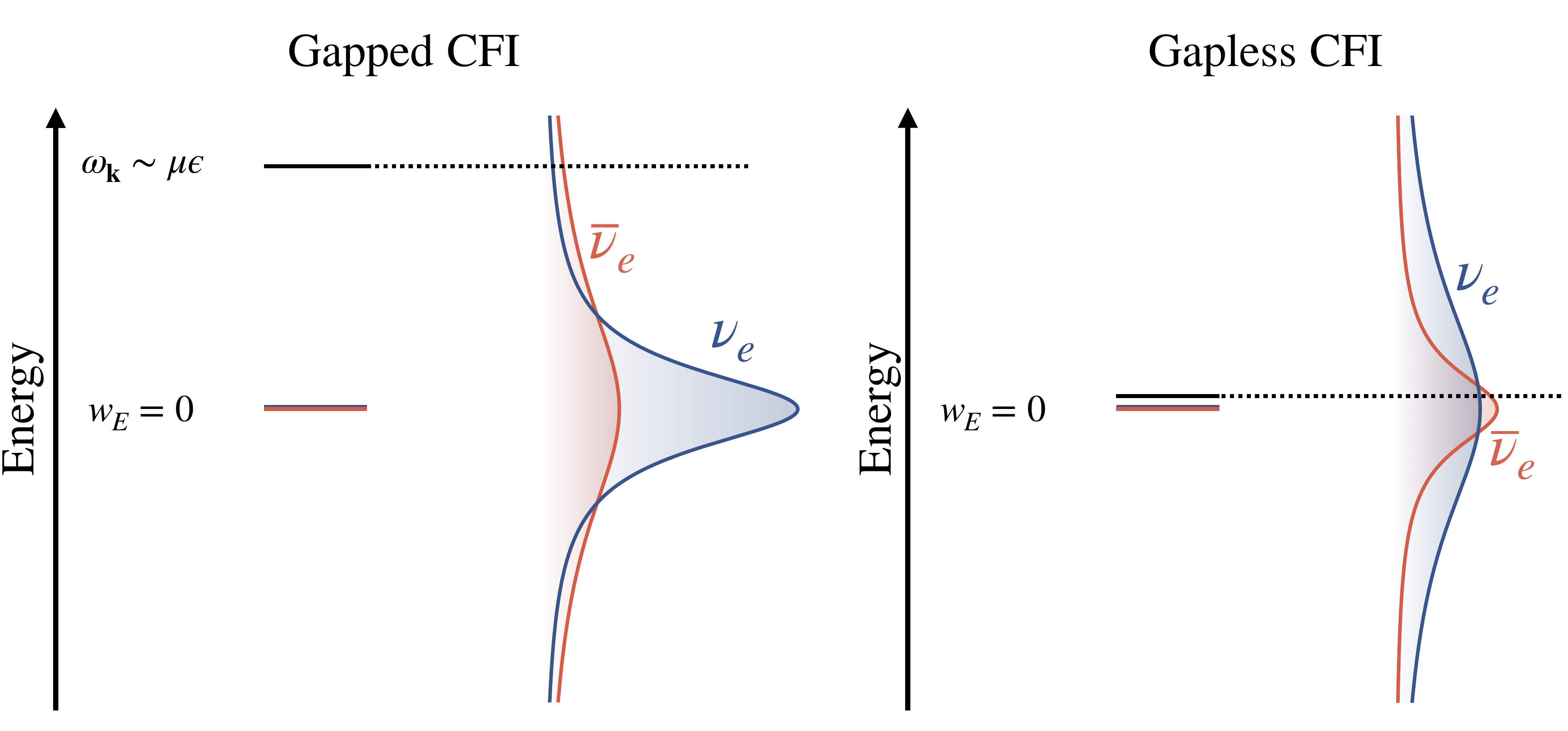}
    \caption{Schematic representation of the CFI mechanism. \textit{(Left)} A subdominant $\overline{\nu}_e$ population with a larger collision rate $\Gamma_{\overline{\nu}_e}>\Gamma_{\nu_e}$ is more broadened in frequency. We show the spectral function of $\nu_e$ and $\overline{\nu}_e$ in color. Thus, flavomons with high frequency (\textit{gapped}) are more efficiently emitted by $\overline{\nu}_e$ than absorbed by $\nu_e$. \textit{(Right)} If \(\Gamma_{\nu_e}>\Gamma_{\bar\nu_e}\), the narrower $\overline{\nu}_e$ distribution can dominate near $w_E=0$, so that flavomons with near-vanishing frequency (\textit{gapless}) are more efficiently emitted than absorbed.}\label{fig:cfi}
\end{figure}

So far we have neglected incoherent neutrino--matter interactions. Throughout the decoupling region of SNe, however, charged-current emission and absorption and neutral-current scattering are unavoidable. Their rates are typically much smaller than those of fast instabilities, whose length scales are far below the neutrino mean free path, but can be comparable to slow-instability growth rates. Slow flavor evolution near decoupling therefore cannot generally be separated from collisions.

Collisions usually damp flavor coherence because they create and absorb neutrinos in flavor eigenstates. However, as pointed out by Johns~\cite{Johns:2021qby}, different interaction rates among flavors can instead produce \textit{collisional flavor instabilities} (CFIs), potentially relevant in dense SN and NSM regions. While CFIs require neutrino--matter collisions, they still drive flavor exchange among neutrinos without incoherent neutrino--neutrino collisions. In this sense, they still fall under the headline of collisionless neutrino flavor instabilities.

Their formulation is subtler than that of fast and slow instabilities. For the latter, without neutrino--matter collisions, a nonthermal stationary background can be assumed: either collective evolution is much faster than collisions, as for fast instabilities, or collisions have become negligible outside the decoupling region, as for slow instabilities. By contrast, an arbitrary nonequilibrium distribution cannot be used as the background of a genuinely collisional instability, because collisions would themselves relax it to thermal and chemical equilibrium, which is necessarily stable by the second law of thermodynamics~\cite{Fiorillo:2025zio}. CFIs therefore require an out-of-equilibrium configuration that cannot relax incoherently: for example, $\nu_e$ and $\overline{\nu}_e$ may be collisionally coupled while the heavier flavors are not, so that the latter can equilibrate only through coherent flavor conversion. Alternatively, the CFI may grow much faster than the collision rate itself, as in the case of resonant CFIs discussed below.

At least part of CFI physics admits the same interpretation in terms of stimulated flavomon emission as collisionless instabilities. Consider an isotropic plasma of $\nu_e$ and $\overline{\nu}_e$, with $n_{\nu_e}>n_{\overline{\nu}_e}$, neglecting vacuum mixing, and focus on the homogeneous scalar mode ($\bK=0$, $e^i=0$). Without collisions, the system supports a stable mode with $\omega_\bk\sim\mu\epsilon$, corresponding to one of the two stable branches in Fig.~\ref{fig:dispersion_stable} at $\bK=0$ (the other branch is a transverse mode with a space-like polarization vector). It is stable because zero-vacuum-frequency neutrinos cannot satisfy the emission condition $\omega_\bk=0$ from Eq.~\ref{eq:virtuality}.

Collisions broaden the neutrino energy levels by $\Gamma_{\nu_e}$ and $\Gamma_{\overline{\nu}_e}$. Therefore, they induce a finite probability for flavomon emission and absorption; in place of the delta function enforcing energy conservation in Eq.~\ref{eq:emission_rate}, we should use the broadened density of energy levels, so that the growth rate will be proportional to
\begin{equation}
    \gamma_\bK\propto
    \frac{n_{\overline{\nu}_e}\Gamma_{\overline{\nu}_e}}
    {\Delta \varepsilon^2+\Gamma_{\overline{\nu}_e}^2}
    -
    \frac{n_{\nu_e}\Gamma_{\nu_e}}
    {\Delta \varepsilon^2+\Gamma_{\nu_e}^2}.
\end{equation}

For the special case of isotropic distribution and $\bK=0$, $\Delta\varepsilon=\omega_\bk-w_E c_V$; in fact, we may simplify the problem even further and take the limit of massless neutrinos $w_E c_V\to 0$, so $\Delta \varepsilon=\omega_\bk$. Flavomons carry negative DLN and are therefore emitted by $\overline{\nu}_e$ and absorbed by $\nu_e$. If $\Delta\varepsilon\gg\Gamma_{\nu_e},\Gamma_{\overline{\nu}_e}$, the mode becomes unstable when
$n_{\overline{\nu}_e}\Gamma_{\overline{\nu}_e}>
n_{\nu_e}\Gamma_{\nu_e}$:
although antineutrinos are less numerous, their broader spectral function can give them a larger off-shell emission probability.

This is the \textit{gapped} CFI of Ref.~\cite{Fiorillo:2025zio}: the mode already exists at zero collision rate and has $\omega_\bk\sim\mu\epsilon\gg\Gamma$. Conversely, if $\nu_e$ interact more rapidly than $\overline{\nu}_e$, the antineutrino spectral function can dominate near $\omega=0$. An instability may then arise from a mode with $\omega_\bk$ of order the collisional widths. Such \textit{gapless} modes exist only because of collisions or nonzero vacuum mixing.

These two schematic mechanisms are shown in the two panels of Fig.~\ref{fig:cfi}. They provide two alternative ways to fulfill the kinematic condition of flavomon emission through the brodening of the levels induced by neutrino--matter collisions. The emergence of these two branches of CFIs was realized in Ref.~\cite{Johns:2021qby} and subsequent works~\cite{Johns:2022yqy,Lin:2022dek,Xiong:2022zqz,Fiorillo:2023ajs,Wang:2025vbx}. The corresponding threshold conditions~\cite{Lin:2022dek,Fiorillo:2025zio} are
\begin{equation}\label{eq:threshold_gapped}
    \int dE\,\frac{dn_{\overline{\nu}_e}}{dE}\,
    \Gamma_{\overline{\nu}_e}(E)>
    \int dE\,\frac{dn_{\nu_e}}{dE}\,\Gamma_{\nu_e}(E),
\end{equation}
for gapped modes, and
\begin{equation}\label{eq:threshold_gapless}
    \int dE\,\frac{dn_{\overline{\nu}_e}}{dE}\,
    \Gamma_{\overline{\nu}_e}(E)^{-1}>
    \int dE\,\frac{dn_{\nu_e}}{dE}\,
    \Gamma_{\nu_e}(E)^{-1},
\end{equation}
for gapless ones. Both follow immediately from the kinematic requirement for flavomon emission we have deduced above. At large frequency the Lorentzian tails scale as $\Gamma/\omega^2$, whose integral over the entire distribution is positive if Eq.~\ref{eq:threshold_gapped}, while at $\omega=0$ their height scales as $\Gamma^{-1}$, yielding the gapless condition when integrated over the distribution.

When the number densities of the dominant and the flipped species approach each other, the excess of flipped neutrinos is as large as it can be. The flavomon width becomes much larger than the collision rate itself; the physics of this phenomenon is rather similar to slow instabilities in the regime of very small total DLN, where, as we have seen, the growth rate scales as $\gamma_\bK\propto \sqrt{w_E \mu}$. Similarly, when $\Gamma_{\nu_\alpha}\gtrsim \mu \epsilon^2$, the growth rate of CFIs also largely exceeds the collision rate and becomes $\gamma_\bK\sim \sqrt{\Gamma_E \mu}$. This enhancement is usually referred to as resonant CFI~\cite{Xiong:2022zqz,Liu:2023pjw}.

This discussion covers only the onset of CFIs. There is of course considerable interest in their nonlinear evolution~\cite{Xiong:2022vsy,Kato:2023cig,Zaizen:2025ptx,Froustey:2025nbi}, also in competition with other forms of flavor instabilities~\cite{Takahashi:2026vci}. At present, the numerical progress in these works has not yet led to a clear theoretical understanding of the relaxation mechanism.

\section{Neutrino--flavomon kinetics}\label{sec:flavomon_kinetics}

Flavomons provide the relevant degree of freedom to understand the onset of instability. As we have seen, by understanding the kinetics of their emission, one can get a qualitative feeling of what causes an instability. After their production in an instability, however, flavomons also play a role in its subsequent relaxation. They propagate, drift, and, through their emission and absorption, feed back on the neutrino distribution. We now review what is currently known about these phenomena.

\subsection{Flavomon propagation in inhomogeneous environments}\label{sec:flavomon_propagation}

While the dispersion relation considers plane-wave states, flavor waves are of course always produced in wavepackets, with a roughly defined position $\br$ and momentum $\bK$. By the uncertainty principle, we have the lower bound $\Delta \br\gtrsim \Delta \bK^{-1}\sim (\mu\epsilon)^{-1}$. Since $\mu\epsilon\sim \mathrm{cm}^{-1}$ while astrophysical environments are typically inhomogeneous over length scales of hundreds of meters or more (although the presence of turbulent fluctuations at smaller scales may affect this argument), a flavomon wavepacket has usually a reasonably well-defined position and momentum.

Inhomogeneous environments cause these to evolve, as they exert a force on flavomons changing their wavevectors, much the same as potential gradients pushing on particles. Thus, the flavomon wavelength can be compressed or widened as it passes through different regions, just as light  passing from air to water.

The first theoretical notice of these propagation phenomena is in Ref.~\cite{Fiorillo:2025ank}, which points out that the propagation of flavor waves with group velocity
\begin{equation}
    \frac{d\br}{dt}=\frac{\partial \Omega_\bK}{\partial \bK}
\end{equation}
could generally induce nonlocal effects in inhomogeneous environments. 
Subsequently, Ref.~\cite{Johns:2025yxa} has advocated more generally for a geometrical-optics treatment of flavor waves, writing down formally the WKB equations which include also the momentum evolution
\begin{equation}
    \frac{d\bK}{dt}=-\frac{\partial \Omega_\bK}{\partial \br}.
\end{equation}
In this section, we always use unless specified $\Omega_\bK$ to refer only to the real part of the eigenfrequency; the imaginary part describes the slow growth of the amplitude along the ray traversed by the flavomon.
These equations are of course the well-known Hamilton equations for a quasiparticle with Hamiltonian $\Omega_\bK$ at a formal level. Their practical implications however are a different story. Recently, Ref.~\cite{Fiorillo:2026tee} has moved in this direction, noting that inside a SN core, below the shock wave, the dominant term in the Hamiltonian $\Omega_\bK=-\lambda(\br)-\sqrt{2}G_F D^0(\br)+\omega_{\bk}(\br)$ (we neglect here the nonrelativistic velocity of matter) is the matter refraction $\lambda(\br)$. Therefore, the quasi-classical equations for flavomon motion can be considerably simplified
\begin{equation}\label{eq:WKB_equations}
    \frac{d\br}{dt}=\frac{\partial \omega_\bk}{\partial \bk},\;\;\frac{d\bK}{dt}\simeq\frac{\partial \lambda}{\partial \br}.
\end{equation}
Weakly unstable flavomons, as we have seen, usually move with a group velocity close to the speed of light, both for fast and slow instabilities. On the other hand, the wavevector of the flavomon can be changed considerably by matter inhomogeneities, which exert a force in the direction of the gradient. Hence, flavomons are refracted towards large matter densities.

As a consequence, if the range of unstable wavenumbers is too narrow, the flavomon mode can be pushed out of instability before it grows nonlinear. Using the equations of motion we can easily deduce the qualitative condition for this to happen. A given mode grows nonlinear over the growth time $\tau \sim \gamma_\bK^{-1}$. Within this time, the change in its wavevector from the force exerted by the matter gradient $\partial \lambda/\partial \br\sim \lambda/\ell$, where $\ell$ is the inhomogeneity length scale, is
\begin{equation}
    \Delta K\sim \frac{\lambda \tau}{\ell}\sim \frac{\lambda}{\ell \gamma_\bK}.
\end{equation}
If $\Delta K$ is comparable with the range of unstable wavenumbers, the instability may simply be prevented from growing, because the mode passes swiftly through the instability range without accumulating enough growth. A similar argument, not based on the flavomon picture but rather on an exact solution in a constant-matter-gradient, two-beam neutrino model, was first presented in Ref.~\cite{Bhattacharyya:2025gds} for fast instabilities. (See however Ref.~\cite{Sigl:2021tmj} for early numerical simulations of fast evolution in matter gradients, which also concluded that the latter can suppress or delay the instability.) On the other hand, fast instabilities are not generally endangered in a strong way by this effect, unless they come from an extremely shallow angular crossing; the concrete condition for weak instabilities is derived in Ref.~\cite{Fiorillo:2026tee}.

Instead, this effect is crucial for slow instabilities below the SN shock wave. The key difference is that in this case $\gamma_\bK\sim w_E$, comparable with the vacuum frequency and therefore much smaller than in the case of fast instabilities. Ref.~\cite{Fiorillo:2026tee} shows, by the same token, that the effect cannot be neglected below the SN shock wave, while it is generally subdominant outside of it, where the density is more rarefied. Below the shock wave, the situation entails as much complication as it can: slow instabilities are altered by the matter gradients, yet they can still accumulate a number of e-folds, defined as $\int \gamma_\bK (t) dt$, comparable to or larger than unity. This means that we can neither safely discard them, nor discard the inhomogeneity effect. A new form of linear evolution is required, able to deal with the evolution of unstable modes in inhomogeneous environments; for this reason, Ref.~\cite{Fiorillo:2026tee} deduces, in addition to the WKB Eqs.~\ref{eq:WKB_equations}, also the additional equations for the evolution of the field amplitude along the rays, ultimately expressing the conservation of the occupation number of flavomons. A direct solution of these equations remains to be achieved, and may offer additional challenges, e.g.\ the passage through exceptional points where multiple branches of modes cross together, a difficulty emphasized by Johns and Kost~\cite{Johns:2025yxa}.

\subsection{Flavomon feedback on neutrinos: quasi-linear theory}\label{sec:quasi-linear}

\begin{figure*}
    \includegraphics[width=\textwidth]{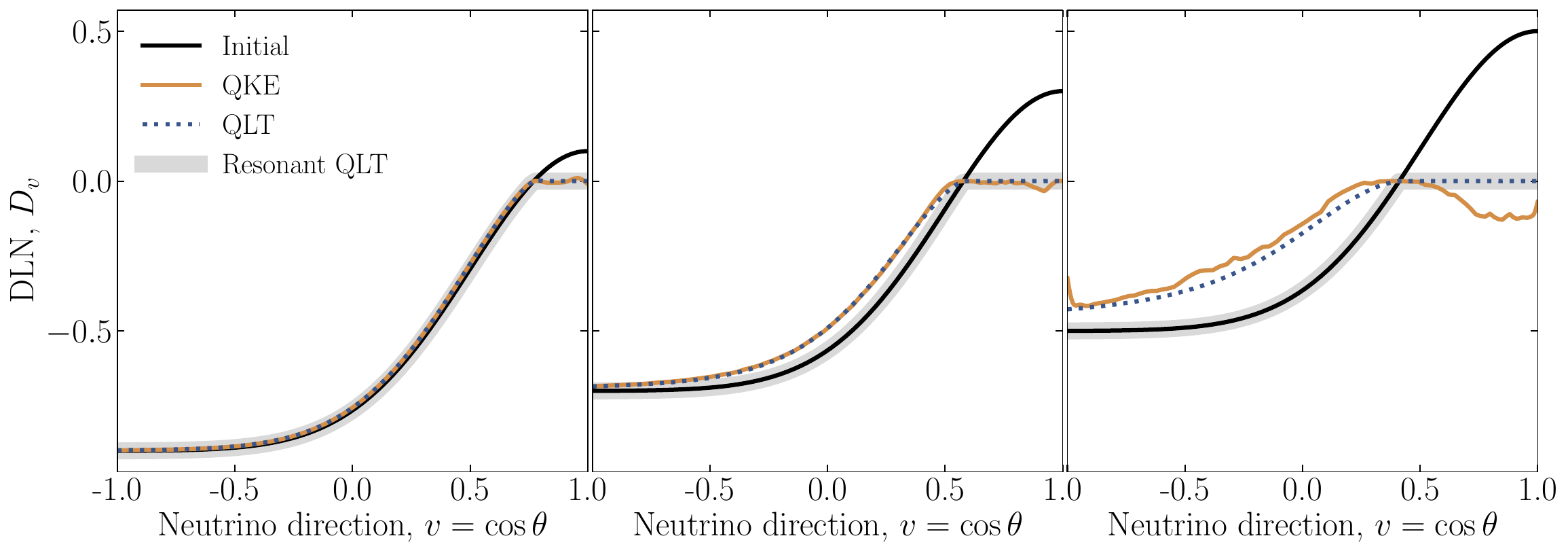}
    \caption{Saturation of fast flavor instability in a homogeneous neutrino background; the initial angular distribution (black) is evolved numerically solving the quantum kinetic equation (QKE) until it reaches a quasi-steady state averaged over time fluctuations (orange). We compare this final state with the predictions of resonant QLT and QLT. Figure adapted from Ref.~\cite{Fiorillo:2026byh}.}\label{fig:qlt}
\end{figure*}

An unstable neutrino distribution does not merely amplify a flavor wave. Every emitted flavomon changes the flavor of the emitting neutrino, depleting the population that drives the instability. The flavomon bath thus feeds back on the neutrino distribution, similar to a laser, where excited atoms decay to their ground state by stimulated emission of photons, amplifying light only up to the point when most of them are not excited. The simplest closure describing the backreaction is given by quasi-linear theory (QLT).

QLT can be understood rather simply by including together with Eq.~\ref{eq:flavomon_emission} the companion kinetic equation for electron and muon neutrinos
\begin{multline}
\partial_t n_{\nu_e,\bp}=-\partial_t n_{\nu_\mu,\bp}
\\
=\sum_{\bK} w(\bK,\bp)
\Bigl[
n_{\nu_\mu,\bp+\bK}
(1-n_{\nu_e,\bp})(1+N_{\bK})
\\
{}-
n_{\nu_e,\bp}
(1-n_{\nu_\mu,\bp+\bK})N_{\bK}
\Bigr].
\label{eq:neutrino_kinetic}
\end{multline}
under the same assumptions as before $|\bK|\ll |\bp|$ and $N_\bK\gg 1$, we obtain

\begin{equation}
    \partial_t n_{\nu_e,\bp}=-\partial_t n_{\nu_\mu,\bp}\simeq -\frac{\Gamma_\bp}{2}(n_{\nu_e,\bp}-n_{\nu_\mu,\bp }),
\end{equation}
with
\begin{equation}
    \Gamma_\bp=2\int \frac{d^3\bK}{(2\pi)^3}w(\bK,\bp) N_\bK.
\end{equation}
The first equality is obvious since the neutrino simply changes flavor in the emission process. Thus, the effect of flavomon emission is to introduce an effective relaxation rate that tends to equalize the two species. This relaxation rate is proportional to the number of flavomons that can kinematically be emitted or absorbed by the neutrino. Overall, this simply describes in equations the Feynman diagram viewpoint: for each flavomon emission event, a flipped neutrino ``unflips'', tending to equalize the two flavors in that mode.

The saturation of weak instabilities is particularly easy to grasp intuitively. Since the probability $w(\bK,\bp)$ is nonzero only for neutrinos exactly on resonance, which are precisely the ones that triggered the instability in the first place, the relaxation consists in the equipartition of the flipped neutrinos that began emitting the flavomons. So in the specific case of weak fast instabilities, where the instability is triggered by an angular crossing with a region of directions dominated by flipped neutrinos, the saturation in this approximation corresponds to equipartition in all of these directions. This conclusion descends directly from the principle of detailed balance applied to the reactions of flavomon emission and absorption.

On the other hand, one easily understands that the equipartition of the flipped side of the crossing cannot be the sole consequence of saturation, as it would violate the conservation of the total DLN---the DLN emitted by flipped neutrinos is stored in the form of flavomons, which are however inaccessible to the averaged neutrino occupation number. In reality, the probability $w(\bK,\bp)$ including the delta function enforcing exact conservation of energy (\textit{resonant approximation}) holds only for infinitely weak instabilities, such that flavomons have precisely defined energy. When $w(\bK,\bp)$ is extended to include the finite width of the flavomon state, the unflipped neutrinos on the other side of the crossing can absorb the flavomons emitted and partially flip. Therefore, the overall relaxation, in addition to the equipartition of the flipped neutrinos, also causes a slight reduction of the unflipped neutrinos, in such a way that the total DLN is conserved.

This general conclusion is ubiquitously found across numerical studies, as reviewed in Sec.~\ref{sec:numerical}. Indeed, QLT has been successfully applied to the relaxation of fast instability in a homogeneous box~\cite{Fiorillo:2026byh}. The results are shown in Fig.~\ref{fig:qlt} in terms of the axially symmetric DLN distribution before and after instability saturation. We show both the outcome of QLT in the resonant approximation---in which the only effect is the equipartition of neutrinos on the flipped side of the crossing---and the full QLT. The comparison with a full numerical solution of the neutrino kinetic equations reveals a striking agreement.

In practice, to perform this calculation beyond the resonant approximation, it is more convenient to start directly from the neutrino kinetic equations. The classical derivation of QLT is based on separating the solution into a slowly varying background and a fluctuating part $\rho_\bp=\langle\rho_\bp\rangle+\delta \rho_\bp$. In the special case that the background is initially flavor-diagonal and homogeneous, $\langle \rho_\bp\rangle$ maintains this form by symmetry at all later times, so it also can be chosen flavor-diagonal and homogeneous, while $\delta\rho_\bp$ encodes the small off-diagonal and inhomogeneous flavor perturbations. QLT assumes that these can be treated as linear at all times, neglecting their nonlinear mutual coupling. Furthermore, the conventional form of QLT as introduced in plasma physics~\cite{vedenov1961stability,vedenov1962quasi,Drummond:1964} works with the squared amplitude of the unstable modes of $\delta \rho_\bp$, which coincides up to the wavefunction renormalization $\mathcal{Z}$ with the occupation number $N_\bK$ of the unstable mode. For this reason, QLT is often said to be based on a random-phase approximation, since it describes the fluctuating response as an ensemble of modes with different $\bK$ whose relative phase is averaged over.

Both of these approximations of QLT need to be critically assessed in each concrete problem at hand. Nonlinear coupling among flavomons is certainly relevant in determining the final-state power spectrum (i.e.\ the final-state $N_\bK$) that they reach upon saturation; the fundamental nonlinear coupling is reported in the Supplemental Material of Ref.~\cite{Fiorillo:2025npi}. A theory of the \textit{weak turbulence} based on this nonlinear coupling among random-phase modes could be constructed, although how to do so in practice is not yet clear. In addition, the assumption that individual modes can be followed separately with an individual occupation number $N_\bK$ may fail at exceptional points where multiple modes become degenerate, a possibility emphasized in Refs.~\cite{Johns:2025yxa,Kost:2026ckc}. For this reason, these works adopt a variant of QLT in which the fluctuating part of the solution $\delta \rho_\bp$ is treated linearly, but it is not expressed in terms of the occupation number $N_\bK$. Therefore, this method is sensitive to the phases of the field; the advantage is of course that the approximation of individually evolving modes with mutual random phases is relieved, but the disadvantage is that the fast phase dynamics of $\delta \rho_\bp$ needs to be followed numerically---in other words, standard QLT only follows amplitude variations over timescales of order $\gamma_\bK^{-1}$, while this method needs to follow phase variations over much shorter timescales of order $\mathrm{Re}(\Omega_\bK)^{-1}$. This may ultimately be necessary because of the exceptional-points difficulty~\cite{Kost:2026ckc}, but a general understanding of this issue is not yet available.

\section{Shifting paradigms in flavor conversions}\label{sec:history}

Our introduction to flavor instabilities is based on the realization that neutrinos form a plasma, characterized by the stimulated emission of collective waves within the plasma which feeds back onto the background neutrino distribution; each emission of a flavomon changes the state of the emitting neutrino. This picture is however very recent, while the viewpoint on flavor conversions has repeatedly changed over the last thirty years. It is therefore useful to track the multiple viewpoints on the subject, and how they all connect to each other. Obviously the list of references can only be incomplete, while we refer to earlier reviews~\cite{Duan:2010bg,Chakraborty:2016yeg,Tamborra:2020cul,Volpe:2023met,Johns:2025mlm,Raffelt:2025wty} providing a more comprehensive description of the older history of the subject.

\subsection[Symmetry-constrained evolution: ``laminar'' flavor conversions]{Symmetry-constrained evolution:\\ ``laminar'' flavor conversions}

The early understanding of flavor conversion in supernova was strictly linked to the symmetries that it was believed to inherit, in particular spherical symmetry and stationarity of the neutrino outflow. For this reason, using a fluid analogy that foreshadows the later developments, we discuss this early phase as focused on the ``laminar motion'' of neutrinos. Even within these symmetry constraints, the primary achievement in this phase was the realization that CFC can develop spontaneously through an instability mechanism. We now briefly review how this realization arose.

While it was realized very early that neutrinos should undergo ordinary MSW refraction~\cite{Wolfenstein:1977ue,Mikheyev:1985zog} in SNe~\cite{Fuller:1987gzx}, the role of neutrino self-interactions in establishing a coherent flavor dynamics was realized first in general, at the level of the kinetic equations~\cite{Notzold:1987ik,Pantaleone:1992eq, Sigl:1993ctk}. The seminal realization by Samuel~\cite{Samuel:1993uw,Kostelecky:1994dt,Samuel:1995ri} that neutrinos can undergo self-induced flavor oscillations, also did not refer to SNe and astrophysical environments, but rather to the early Universe. 

Between 1995 and the early 2000s, it was then realized that neutrino self-refraction does appear in SNe and alters their flavor evolution~\cite{Qian:1994wh,Pantaleone:1994ns,Sigl:1994hc}, although this early phase was primarily focused on the impact that self-refraction has on the standard MSW resonance. It became particularly apparent that self-interactions can \textit{synchronize} the MSW resonance, causing neutrinos with different vacuum frequencies to undergo a common MSW transition~\cite{Pastor:2001iu,Pastor:2002we}. 

The focus shifted abruptly around 2005-2006, with the realization that neutrinos can exhibit collective flavor transformations among themselves at radii much closer to the center than the MSW resonance is approached~\cite{Fuller:2005ae,Duan:2005cp,Duan:2006an,Duan:2006jv,Duan:2007mv}. The laminar nature of evolution was incorporated in the very common ``bulb model'', assuming neutrinos are steadily emitted from an inner surface. 

In the bulb model, the key feature understood in this epoch was the possibility of bipolar oscillations. In modern language, what was understood at the time was that a plasma of $\nu_e$ and $\overline{\nu}_e$ exhibits an instability whereby the two species exchange flavor with one another. This instability is of the same kind as the slow collisionless instability we have discussed above. However, because of the laminar restriction to a perfectly static emission, the early discussions could only find such an instability for $\Omega=0$. Therefore, in this phase collective conversions were believed to be associated only with relatively outer radii, much outside the shock wave of the SN. Because of the restriction to $\Omega=0$, in this setup there is a single unstable mode, whose evolution does not at all resemble the ``turbulent'' behavior of multiple excited flavomon modes. Therefore, the expected evolution in this epoch was rather regular and analogous to that of a pendulum~\cite{Hannestad:2006nj}, with the $\nu_e$ and $\overline{\nu}_e$ periodically flipping to $\nu_\mu$ and $\overline{\nu}_\mu$. Ultimately, the reason for this beautiful pendulum analogy is the existence, for $\Omega=0$, of a host of conserved quantities~\cite{Raffelt:2011yb,Pehlivan:2011hp}. One should also stress that, while the bipolar instability is ``slow'' in the sense of being induced by the vacuum neutrino masses, the restriction to $\Omega=0$ caused it to emerge only at such large radii that the neutrino density was significantly reduced. The growth rate of this bipolar instability was consistently found to be of the order of $\gamma\sim \sqrt{w_E \mu}$, which was taken to be a defining feature of slow instabilities, even though, as we now know, it applies only to the regime of small neutrino-antineutrino asymmetries $w_E\gg \mu \epsilon^2$. Thus, the bipolar instability belongs to that class of slow instabilities that are, so to speak, faster than the fast instability, since $\sqrt{w_E \mu}\gg \mu \epsilon$. The bipolar instability itself is of the same kind as that originally discussed by Samuel many years before in the context of the early Universe~\cite{Samuel:1995ri}.

Thus in this phase two forms of laminar collective motion had been identified: the synchronized precession, where neutrinos with all energies evolve in flavor maintaining relative coherence (formally their $\psi_\bp$ maintains a common phase for all energies), and the bipolar instability, where two sectors, e.g.\ the $\nu_e$ and $\overline{\nu}_e$ population, periodically exchange flavor with each other, e.g.\ through the $\nu_e\overline{\nu}_e\to \nu_\mu\overline{\nu}_\mu$ coherent transformation. The phenomenological hallmark of these conversions was looked at in the form of the so-called spectral splits. As the neutrinos move out through the SN progenitor, and the density becomes progressively lower, they stick adiabatically to the coprecessing or bipolar solution. Due to the assumed adiabaticity, they finally leave as mass eigenstates; the constraint of lepton number conservation then enforces the emergence of sharp splits in the flavor composition~\cite{Raffelt:2007cb,Raffelt:2007xt}, which were observed numerically in simulations of the bulb model~\cite{Duan:2007fw,Duan:2007bt}. The split phenomenology gained increasing prominence and was extended to three-flavor studies~\cite{Dasgupta:2009mg,Duan:2008za}. Meanwhile, the first concerns about the impact of a multi-angle distribution for neutrinos began to appear\hbox{~\cite{Esteban-Pretel:2007jwl,Esteban-Pretel:2008ovd,Chakraborty:2011gd,Cherry:2012zw}.}

\subsection[Spontaneous symmetry breaking: ``turbulent'' flavor conversions]{Spontaneous symmetry breaking:\\ ``turbulent'' flavor conversions}

The subsequent developments led more or less to a collapse of the old symmetric, static vision of the neutrino outflow. This happened gradually with the realization that several kinds of instabilities, spontaneously breaking the symmetries imposed on the problem, may emerge, first in the form of static, multi-angle instabilities~\cite{Banerjee:2011fj,Raffelt:2013rqa,Duan:2014gfa,Mangano:2014zda,Abbar:2015mca,Chakraborty:2015tfa}, and, finally, in the form of temporal instabilities~\cite{Abbar:2015fwa,Dasgupta:2015iia}. In short, this marked the transition from laminar evolution of neutrinos with specified symmetries, to ``turbulent'' evolution. A laminar, steady solution to the Navier-Stokes equations becomes unstable to time-dependent perturbations above a critical Reynolds number, even if the external conditions are static. In the same way, the static bulb-model evolution is unstable to time-dependent perturbations, so that neutrino flavor evolution spontaneously breaks time translation invariance, as well as all the other symmetries of the medium.

\subsubsection{Fast instabilities}

In parallel, the community progressively realized the existence of fast instabilities even at zero vacuum frequency. The first remarks in this direction actually came much earlier, with a series of papers by Sawyer between 2004 and 2009~\cite{Sawyer:2004ai,Sawyer:2005jk,Sawyer:2008zs} showing for discrete neutrino beams the emergence of fast instabilities. Sawyer later realized that such instabilities may occur immediately outside the SN core~\cite{Sawyer:2015dsa}, a realization that captured the attention of the community~\cite{Chakraborty:2016lct,Dasgupta:2016dbv}, and finally led to the modern approach of solving the dispersion relation for generic space-time dependent modes~\cite{Izaguirre:2016gsx}.

So between 2012 and 2016 three groundbreaking realizations intertwined: that massless neutrinos can exhibit instabilities, that instabilities can spontaneously break the symmetries of the background, and that they can occur close to or even within the protoneutron star at the center of the SN, where neutrinos have substantial feedback on the surrounding matter. For this reason, describing collective flavor evolution transitioned from being a particle-physics problem, with a potential impact on the neutrino signal from a galactic SN, to an astrophysical problem, with potential impact on the SN evolution as a whole.

At the conceptual level, the subsequent research focused on outlining more clearly the conditions for an instability. Several works have focused on the general properties of the dispersion relation of fast instabilities, attempting a general classification of its solutions~\cite{Capozzi:2017gqd,Airen:2018nvp,Yi:2019hrp,Capozzi:2019lso}. The mathematical arguments due to Morinaga~\cite{Morinaga:2021vmc} showed that a fast flavor instability appears whenever the energy-integrated angular distribution of neutrinos exhibits a crossing, i.e.\ a transition from $\nu_e$-dominance to $\nu_\mu$-dominance. A crossing is in fact required for an instability of any kind, as later shown in Ref.~\cite{Dasgupta:2021gfs}, although the physical argument based on lepton number conservation we gave in Sec.~\ref{sec:stable_plasma} was formulated only later, for fast instabilities by~\cite{Johns:2024esf}, and in general in~\cite{Fiorillo:2024bzm}. The physical proof of Morinaga's theorem, based on the kinematic conditions for Cherenkov emission of flavor waves, was also given in~\cite{Fiorillo:2024bzm}.

\subsubsection{Theoretical paradigms}

Within the fast instability paradigm, the very formulation of the problem of flavor conversions had to change. One no longer considers a boundary emitting neutrinos in unstable states, since the boundary itself (the protoneutron star) is affected by the neutrinos undergoing flavor conversions. Even more importantly, the presence of modes growing within a length scale much shorter than the astrophysical one makes the notion of a sharp emitting surface basically untenable. Therefore, the very formulation of the flavor conversion problem has come to be questioned; after all, the appearance of an instability is in itself a signal that flavor evolution has not been correctly followed, since flavor evolution reacts precisely to the instability. Therefore, most approaches in the community are based on the picture that small-scale flavor waves sourced by an instability, as soon as the latter emerges, feed back onto the neutrino distribution.

In response to the fast-instability problem, many approaches have followed the paradigm of local, or sub-grid, relaxation, assuming that neutrinos relax in each volume element to a local equilibrium state. The nature of this local equilibrium is of course a central challenge to address, with several works adopting equipartition within the constraints of conservation laws (e.g.\ Ref.~\cite{Wu:2017drk,Li:2021vqj,Just:2022flt,Ehring:2023lcd,Ehring:2023abs} for first practical implementations of this assumption in NSM and SN simulations); a more detailed discussion is in Sec.~\ref{sec:numerical}. One should stress that the notion of sub-grid relaxation, while appearing very natural, is still mostly postulated and not proved; this is especially true for slow instabilities, which grow over length scales potentially comparable with the astrophysical ones, hinting at nonlocal relaxation.

The local-equilibrium assumption was subsequently incorporated and developed by Johns into a framework termed \textit{miscidynamics}~\cite{Johns:2023jjt,Nagakura:2023jfi,Johns:2024dbe}. In its present formulation, its defining condition is that the coarse-grained neutrino distribution $\langle\rho_\bp\rangle$, averaged over spatial scales comparable with the wavelength of collective modes, remains close at every point to a local mixing equilibrium, meaning that it satisfies $\left[\langle{\sf\Omega}_\bp\rangle,\langle\rho_\bp\rangle\right]=0$, where $\langle {\sf\Omega}_\bp\rangle$ is the coarse-grained neutrino self-energy. As stressed by Johns, this state need not be flavor-diagonal. This condition is analogous to, although somewhat more general than, the sub-grid relaxation, in that it also assumes local relaxation. On the other hand, there are infinitely many possible coarse-grained distributions satisfying it: a predictive theory requires of course also a concrete specification of the equilibrium achieved. More recent formulations explicitly separate miscidynamics from the thermodynamic and ergodic hypotheses~\cite{Johns:2025yxa}; once these hypotheses are relinquished, however, the local-equilibrium condition must be supplemented by some other dynamical closure to predict the relaxed distribution.

Miscidynamics originally proposed this closure through the hypothesis of ergodicity and constrained maximization of a coarse-grained entropy~\cite{Johns:2023jjt}. This was a substantive and testable hypothesis, attempting to select a particular state from the manifold of stationary configurations. The coarse-grained neutrino entropy does indeed grow in flavor conversions, as shown in Ref.~\cite{Fiorillo:2025zio} in the framework of QLT; however, this growth does not imply evolution toward a maximum-entropy state. Furthermore, Ref.~\cite{Fiorillo:2025kko} exhibited configurations in which the usual conservation laws would allow flavor conversion, whereas the kinematics of flavomon emission forbids the corresponding instability. Therefore the state dynamically accessible to the system cannot in general be inferred solely from conservation laws and entropy maximization.

Meanwhile, Refs.~\cite{Fiorillo:2024fnl,Fiorillo:2024qbl,Fiorillo:2024bzm,Fiorillo:2024uki,Fiorillo:2024dik,Fiorillo:2024pns,Fiorillo:2025ank,Fiorillo:2025npi,Fiorillo:2025zio,Fiorillo:2025kko} have developed the theory of the neutrino plasma in the form summarized in Secs.~\ref{sec:theory_plasma} and~\ref{sec:instabilities}. The core of this theory is that flavor waves, and their quanta, the flavomons, can be treated as independent degrees of freedom---the emergent quasiparticles of the neutrino plasma---whose emission and absorption explains the flavor instabilities deduced from the conventional dispersion relation. This is not merely a change in vocabulary: beyond giving an intuitive understanding of what flavor instabilities are, this framework provides an explicit computational framework to deduce the growth rates of weakly unstable flavomon modes through the Feynman diagrammatics, and, most importantly, it incorporates the feedback on the neutrino distribution, at least approximately, through the quasi-linear method.  

Recently, Johns and Kost~\cite{Johns:2025yxa} have developed a vision of local flavor equilibrium that is based on analogous concepts. Perturbations around a coarse-grained background are decomposed into normal modes that propagate, grow or damp, analogous to flavomons; feed back on the averaged neutrino distribution, analogous to quasi-linear theory; and interact through higher-order wave--wave couplings, analogous to the mutual flavomon couplings introduced in Ref.~\cite{Fiorillo:2025npi}. While this novel framework is often juxtaposed with miscidynamics, its founding concepts---flavor waves as independent degrees of freedom, similarly to the neutrino plasma theory---are independent of the local equilibrium hypothesis, especially since flavor waves are propagating degrees of freedom. Of course, in order for flavor waves to be defined as linear excitations, the background itself should be an equilibrium, i.e. it should satisfy the local equilibrium postulate of miscidynamics. 

\subsubsection{Multiple forms of instability}

Another aspect that has been tackled by the recent literature is the relative impact of different forms of flavor instabilities. The realization that fast instabilities can occur over timescales of picoseconds, way below any timescale associated with neutrino masses, led to a general excitement over the evolution in the limit of negligible neutrino masses, so that many numerical works focused on this limit. A general warning against this approximation came from the numerical works of the Copenhagen group~\cite{Shalgar:2020xns,DedinNeto:2023ykt}; see Sec.~\ref{sec:numerical} for more details. A general theory of slow flavor instabilities in Refs.~\cite{Fiorillo:2024pns,Fiorillo:2025ank} shows explicitly that neutrino distributions which are fast-stable may become slow-unstable, and in fact for axially symmetric distributions they certainly are if they have a single spectral crossing~\cite{Fiorillo:2025kko}. Since angular crossings emerge only relatively late in the SN evolution, slow instabilities are the first to appear in SNe~\cite{Fiorillo:2025gkw}.

A separate branch of instability relates to the impact of collisions. The seminal work by Johns~\cite{Johns:2021qby} introduced the idea that a mismatch in the collision rates of $\nu_e$ and $\overline{\nu}_e$ can induce the novel branch of CFIs. Several later works have generalized the original discovery to anisotropic, single-energy models~\cite{Johns:2022yqy}, single-angle distributions with continuous energy spectra~\cite{Lin:2022dek,Xiong:2022zqz,Wang:2025vbx}, and generically inhomogeneous modes~\cite{Fiorillo:2025zio,Liu:2026roe}. One important aspect for SNe is that a CFI must be of the gapless nature (while in NSMs gapped instabilities are more common, e.g. Ref.~\cite{Froustey:2026vhm}), since the flipped species $\overline{\nu}_e$ is the least interacting one (see Fig.~\ref{fig:cfi}), and so it can only appear under the condition in Eq.~\ref{eq:threshold_gapless}. Since in SNe this condition is only marginally satisfied, such instabilities are generally expected to be very slow~\cite{Wang:2025vbx, Fiorillo:2025gkw}. Interestingly, the condition for this to happen can be written very simply in order of magnitude; since for beta processes $\Gamma_{\nu_e}/\Gamma_{\overline{\nu}_e}\sim n_n/n_p$, where $n_n$ and $n_p$ are the neutron and proton number densities, approximately we expect collisional instabilities above threshold for $n_{\overline{\nu}_e}\gtrsim n_{\nu_e} n_p/n_n$~\cite{Fiorillo:2025gkw}.

\subsection{Exact solutions: flavor pendula and solitons}

At the theoretical level, the general interest in exact, laminar solutions of the neutrino kinetic equations has persisted in the recent period. Even though such regular evolution may play no role in realistic environments, it nevertheless provides an important baseline for understanding nonlinear flavor evolution.

Even for fast instabilities, it was quickly realized that the restriction to homogeneity ($K=0$) leads to a regular, pendular-like behavior~\cite{Johns:2019izj,Padilla-Gay:2021haz} for an arbitrary angular distribution. This regularity was speculated to come from the existence of conserved quantities, just as in the case of the slow flavor pendulum. The reality of this statement was understood in Ref.~\cite{Fiorillo:2023mze}, which showed that in fact the fast flavor pendulum can be directly mapped to the slow flavor pendulum and possess the same family of invariants of motion called Gaudin invariants. Ref.~\cite{Xiong:2023upa} provided a geometric view of the integrability, tracing it to a symmetry of the unstable mode. Generally, breaking this symmetry, or the associated conservation law, in any way, destroys the beautiful periodicity of the solution~\cite{Fiorillo:2023hlk,Xiong:2023upa}.

In fact, these time-dependent solutions are special members of a more general class of space-time-dependent exact solutions known as flavor solitons~\cite{Fiorillo:2023hlk}, which play an analogous role to solitons in electronic plasmas. The stability condition for these exact solutions can be investigated on the basis of the well-known Nyquist criterion~\cite{Fiorillo:2023hlk,Fiorillo:2024dik}, which provides another tie to the physics of electromagnetic plasmas. On the other hand, when the restriction to a single unstable mode is removed, the extension of the Nyquist criterion, adopted in Ref.~\cite{Dasgupta:2025quc}, provides only a formal guidance, since one would need to know beforehand the polarization of the unstable eigenmode to apply it.

The regular behavior in the presence of neutrino-matter collisions has also been investigated, with Ref.~\cite{Padilla-Gay:2022wck} observing a progressive damping of the flavor pendulum due to collisions, provided that the conditions for a collisional instability are not met. In these perfectly homogeneous setups, a general method of solution based on the separation of the fast self-interaction-induced oscillations and the slow collisional evolution was given in Ref.~\cite{Fiorillo:2023ajs}.

Recently, Ref.~\cite{Liu:2025muc} has uncovered a new exact solution that again follows a pendular behavior without requiring a homogeneous mode; this is based on two neutrino beams in the fast regime excited with a single plane-wave perturbation. Exact solutions with a nonzero wavenumber without a temporal instability had earlier been introduced~\cite{Duan:2021woc}, but Ref.~\cite{Liu:2025muc} extends them to unstable modes. The regularity here does not follow from homogeneity, but from the constraint of only two massless neutrino beams with a single unstable wavenumber. As shown in Ref.~\cite{Fiorillo:2026jgw}, the regularity then follows from the restricted number of degrees of freedom; this descends from a general statement~\cite{Fiorillo:2026lyz} that the dynamics of three mutually precessing spins with a conserved one is bound to be pendular. The mutual connection among all these exact solutions has been recently reviewed in Ref.~\cite{Fiorillo:2026jgw}.

While all the above examples possess regular dynamics by virtue of artificial symmetries or restrictions, it is interesting that even a realistic unstable neutrino plasma may exhibit regular dynamics in a transient phase when the instability is very weak. The reason is that in this regime the range of unstable wavenumbers may be so narrow as to behave like a single mode. Therefore, the neutrino distribution, rather than feeling a feedback from an ensemble of modes with random mutual phases, interacts coherently with one mode---in the flavomon language, it emits coherently flavomons with a single wavenumber. The process can therefore remain reversible. This is shown explicitly in Ref.~\cite{Fiorillo:2026att}: a weakly unstable plasma shows cyclic evolution where the flipped neutrinos $\overline{\nu}_e$, rather than simply equipartitioning and removing the crossing as expected from QLT, convert completely to $\overline{\nu}_\mu$ and then back periodically. This behavior is again described by a pendular dynamics, and was already found in electronic plasmas by O'Neil and collaborators~\cite{o1971nonlinear,o1972nonlinear}. So extremely weak instabilities fail to follow QLT, because the limited number of excited modes does not behave incoherently as assumed in QLT.

\subsection{Neutrino--neutrino correlations beyond the mean field}

A question that has received more and more attention is whether the neutrino mean-field approximation, i.e.\ the treatment of their mutual interaction as a mean potential term in their effective Hamiltonian, is applicable. The debate goes back to Ref.~\cite{Bell:2003mg} claiming a strong speedup from the development of entanglement among the particles. This possibility was analyzed in detail by Friedland and Lunardini~\cite{Friedland:2003dv,Friedland:2003eh}, who concluded that in the limit of many particles the mean-field evolution agrees with the many-body evolution within its range of applicability. One should stress that the mean-field equations were derived in Ref.~\cite{Sigl:1993ctk} through a controlled approximation in the Fermi constant $G_F$, and one would therefore expect it to be valid over the coherent timescale $t_{\rm coh}\sim (G_F n_\nu)^{-1}$ associated with refractive dynamics. The next order in $G_F$ leads to incoherent processes over the much longer timescale $t_{\rm scat}\sim (G_F^2 n_\nu T^2)^{-1}$, where $T$ is the neutrino temperature; this corresponds to neutrino--neutrino scattering. Friedland and Lunardini~\cite{Friedland:2003dv,Friedland:2003eh} did not find a breaking of this hierarchy.

However, the emergence of instabilities has led to a part of the community questioning again the mean-field theory (see e.g.\ Refs.~\cite{Birol:2018qhx,Cervia:2019res,Roggero:2022hpy,Siwach:2022xhx,Siwach:2024jet}). The general approach of these works is to consider spin Hamiltonians derived from including only the forward-scattering among neutrinos, and comparing exact solutions of such Hamiltonians with their mean-field evolution. This approach has been generally criticized~\cite{Johns:2023ewj,Shalgar:2023ooi} because the forward-scattering Hamiltonian is not the true Hamiltonian. Moreover, the all-to-all structure of the forward-scattering spin Hamiltonian may artificially maximize entanglement: every momentum mode interacts repeatedly with every other mode. Instead, in the real neutrino--neutrino Hamiltonian, two neutrino wavepackets interact only very rarely with each other. This last feature has been incorporated in a phenomenological model called once-in-a-lifetime~\cite{Kost:2024esc,Kost:2025vyt}, in which each time a neutrino wavepacket interacts with the background it becomes entangled with a single partner, after which the two particles are assumed never to interact again. In agreement with the expected hierarchy, the resulting beyond-mean-field corrections accumulate on the collisional timescale, being suppressed by an additional power of $G_F$ relative to the coherent mean-field evolution.

Recent numerical studies have attempted to go beyond the forward-only approximation by retaining momentum-changing processes in the exact many-body evolution~\cite{Cirigliano:2024pnm,Froustey:2026slw}. This is an important step, but its present interpretation requires caution. The Boltzmann description is a kinetic limit: it relies on a continuum of available states, a macroscopic number of particles, and the effective loss of correlations embodied in molecular chaos. Current calculations are rather far from that limit, containing up to ten neutrinos and about twenty momentum modes. It therefore remains open whether any observed discrepancy will survive the passage to the macroscopic limit. 

\section{Numerical progress in CFCs}\label{sec:numerical}

After the realization that fast instabilities result in the formation of flavor inhomogeneities with very short length scales, and within a timescale much shorter than the hydrodynamical timescale, a vast numerical effort has been devoted to characterizing this state and implementing its consequences in realistic astrophysical environments. The resulting literature is quite huge, so we can only provide a very coarse view of the subject. Nevertheless, this literature is absolutely essential to understanding the behavior of neutrino plasmas, since it provides the controlled experiments through which one can build intuition about the phenomena characterizing this system.

\subsection{Local approaches}

Given that fast instabilities occur so rapidly, one promising approach is to consider their relaxation in small volume elements (or ``boxes'') of the order of a few centimeters, the wavelength of the unstable modes. The resolution requirements are not so stringent, so numerical solutions are widely available. This approach was initiated more or less simultaneously by many works~\cite{Martin:2019gxb,Capozzi:2017gqd,Bhattacharyya:2020jpj,Bhattacharyya:2022eed}, increasingly evolved into a general effort to identify the final outcome of local fast flavor instabilities~\cite{Bhattacharyya:2022eed,Richers:2021nbx,Wu:2021uvt,Richers:2021xtf,Cornelius:2023eop}, leading to a summary code comparison paper~\cite{Richers:2022bkd}. One of the most important conclusions was the removal of the angular crossing~\cite{Wu:2021uvt}, explained by detailed balance of flavomon emission and absorption~\cite{Fiorillo:2024qbl,Fiorillo:2025npi,Fiorillo:2026byh} as explained in Sec.~\ref{sec:quasi-linear}.

Recently, this strategy has been developed into a promising approach for implementation in realistic environments through the development of sub-grid recipes, i.e.\ mapping the initial unstable configuration to a final quasi-steady configuration fitted to the numerical simulations. This approach was pioneered in Refs.~\cite{Zaizen:2022cik,Zaizen:2023ihz}, and has since been extended and validated by other groups, also through neural network approaches, to predict the final state~\cite{Xiong:2023vcm,Abbar:2023ltx,Abbar:2024ynh,George:2024zxz,Richers:2024zit,Goimil-Garcia:2025ozm,Goimil-Garcia:2026flm}. These sub-grid recipes have since been applied explicitly to global transport~\cite{Xiong:2024pue,Xiong:2026bmo,Lund:2025jjo,Akaho:2025giw,Wang:2025nii,Wang:2025ihh,Akaho:2026kff}, sometimes termed effective classical transport (ECT)~\cite{Xiong:2026bmo}.

One should stress that the applicability of sub-grid recipes to a global environment is not a priori guaranteed. Even if fast instabilities are rapid compared to the time and length scales of astrophysical environments, they source waves whose later fate may well be nonlocal, especially since weak fast instabilities are convective~\cite{Fiorillo:2025ank}. Similar concerns have been voiced at the numerical level~\cite{Cornelius:2023eop}. On the other hand, first implementations of sub-grid fast flavor relaxation seem to provide reasonable agreement with other global approximations discussed below, based on attenuating the strength of the neutrino--neutrino interaction~\cite{Xiong:2024pue}, although this is at present mostly empirical evidence based on comparing two different approximation schemes. A solid theoretical reason for the validity of any of these approaches would be a welcome addition to the field.

Currently, the main theoretical prediction for the saturated state of fast instabilities comes from QLT~\cite{Fiorillo:2024qbl,Fiorillo:2025npi,Fiorillo:2026byh}, which correctly predicts the removal of the angular crossing observed in numerical simulations without having to follow the rapid phase variations of flavor waves, and, especially, interprets it as detailed-balance equilibrium of flavomon emission/absorption. However, how generic such a prediction would be in a space and time-dependent environment is not at all clear, see the examples in Ref.~\cite{Kost:2026ckc}.

Recently, some first attempts at implementing local relaxation of slow instabilities have also emerged~\cite{Padilla-Gay:2025tko,Goimil-Garcia:2026flm,Chen:2026oym}. One should stress, though, that for slow instabilities the problem of nonlocality is exacerbated, since their growth length scale is comparable, at least at their first appearance, with the hydrodynamical length scale.

\subsection{Global approaches}

Solving the kinetic equation in global environments on hydrodynamical time and length scales is impossible. A constructive approach is to attenuate the Fermi constant, so that the wavelength of flavor waves is increased, making the problem numerically tractable. This approach, pioneered in Ref.~\cite{Nagakura:2022kic}, has since been applied in several astrophysical contexts~\cite{Nagakura:2022xwe,Nagakura:2023mhr,Nagakura:2023wbf,Xiong:2024pue,Urquilla:2026bff}. Recently, Ref.~\cite{Zaizen:2026wvj} has warned against a potential caveat from the artificial attenuation, namely the modification of the hierarchy between the inhomogeneity length scale and the flavomon wavelength. This can alter the rate of drift of flavomons across the unstable wavenumber range, as explained in Sec.~\ref{sec:flavomon_propagation}, and artificially weaken instabilities. For slow instabilities, as usual, this problem is exacerbated~\cite{Fiorillo:2026tee}, due to the occurrence of yet another length scale, the growth length scale.

In parallel to this progress, several works from the Copenhagen group~\cite{Shalgar:2019qwg,Padilla-Gay:2020uxa,Shalgar:2022rjj,Shalgar:2022lvv,Shalgar:2024gjt,Cornelius:2024zsb} have attempted to solve the kinetic equations in global astrophysical environments \textit{without} attenuating the interaction Hamiltonian. The spatial resolution for such calculations is usually far coarser than required to recover the wavelength of most unstable modes from linear stability analysis, yet surprisingly in several cases they report reasonable agreement with attenuated methods in the mean features of the neutrino distribution. These works have highlighted several important features of flavor conversions, such as the relevance of neutrino masses and the potential nonlocal effects of flavor instabilities~\cite{Cornelius:2023eop}, although the use of insufficient numerical resolution has been debated in the literature~\cite{Nagakura:2025brr}. Generally, the reported numerical convergence, in spite of the presence of unresolved structures, is an interesting open question in its own right.

\subsection{Quantum moments}

A challenge to be faced when dealing with flavor transport in realistic astrophysical environments is that even the classical transport, without flavor conversions, is substantially challenging from the computational viewpoint. For this reason, SN and NSM simulations often do not offer a complete neutrino angular distribution, but rather only a few angular moments of that distribution, which are propagated through the classical dynamics based on some closure relations. Based on this, several works have explored whether a purely moment-based technique may inform the flavor evolution. Pioneering works in this respect have studied the cascading of power from low to high angular moments~\cite{Johns:2019izj,Johns:2020qsk}. Several works have focused on diagnosing or evolving flavor instabilities based purely on moments~\cite{Myers:2021hnp,Froustey:2023skf,Froustey:2024sgz,Grohs:2025ajr}. It remains unclear how generic the success of a moment-based strategy is. As discussed in Sec.~\ref{sec:fast_instabilities}, nonresonant instabilities with $\gamma\gtrsim \Omega_R$ can be well-represented by the coarse structure of the angular distribution, since all neutrinos are able to participate in the emission of the largely off-shell flavomons. Similarly, for $k=0$, the resonant denominator in the flavor dielectric function becomes independent of the angle, so that moment-based techniques can again be efficient. However, for weak instabilities, as we have seen, the amplitude for flavomon emission depends rather sensitively on the angular distribution along the directions that are kinematically able to radiate flavor waves. For slow instabilities at their first appearance this is a particularly troublesome point, since flavor waves are radiated primarily by neutrinos collinear with them. This suggests that the fine-grained angular information may ultimately be required at least at the first appearance of flavor instabilities.

\subsection{Boundary and initial conditions} 

In parallel with these approaches, it is becoming ever more apparent that the specific boundary conditions, as well as the conditions under which the background changes, can directly impact the outcome of flavor conversions. For example, Zaizen and Nagakura~\cite{Zaizen:2023wht} pointed out that, in a problem with boundary conditions with nontrivial spatial structure rather than initial conditions, the often-found removal of the angular crossing does not in fact occur. They consider two emitting surfaces in front of one another, motivated by previous observations of a swap phenomenon in Ref.~\cite{Nagakura:2023wbf} and by a setup considered in unpublished numerical work by Manu George~\cite{Wu_priv}.

They consider two emitting surfaces in front of one another, so the overall boundary is highly nonlocal. In this case, the boundary condition enforces a complete flavor swap. While this occurs in a toy model, similar features seem to have been observed more recently in realistic environments without externally prescribed boundaries~\cite{Xiong:2026bmo}. Another issue is the approach to instability itself; while most treatments consider the evolution of an unstable configuration, in reality what likely occurs is that a system is slowly driven, through neutrino emission and absorption processes from matter, into an unstable configuration. Ref.~\cite{Fiorillo:2024qbl} shows that the system then remains in the simplest conditions at the edge of instability, removing the angular crossing that triggers the instability in the first place; but the evolution can become much less trivial when multiple configurations with marginal stability cross. Later works~\cite{Liu:2024nku,Urquilla:2025idk,Kost:2026ckc} have expanded on this notion, systematically investigating the potential breakdown of the asymptotic-state paradigm when the background is spatially or temporally varying.

A separate chapter regards the impact that flavor conversions have on the astrophysical evolution of compact transients. Since this topic is somewhat more related to astrophysics, rather than to the theory of the neutrino plasma of interest to nuclear physics, we refer the reader to several comprehensive reviews that have recently appeared on the topic~\cite{Tamborra:2020cul,Volpe:2023met,Johns:2025mlm,Raffelt:2025wty}.

\section{Outlook}\label{sec:outlook}

Understanding the collective flavor evolution of neutrinos has long been recognized as a challenging problem. The nonlinearity of the equations, the vast hierarchy of scales, and the simultaneous dependence on time, space, momentum, and flavor all contribute to this difficulty. Yet ultimately these technical complications  may not be its deepest source. The equations governing neutrino kinetics can be written rather compactly; what has remained hard is to absorb their qualitative content.

For much of its history, the challenge has been approached through the language of oscillations: individual neutrinos oscillate in a refractive potential generated self-consistently by all the others. This description is formally legitimate, just as a plasma can be described by following every electron through the electromagnetic field produced by all the others. It is nevertheless not the language in which its dynamics ultimately becomes intelligible. Instead, it is the collective degree of freedom, i.e.\ the \textit{field}, that comes to the forefront as the variable whose evolution can be understood.

This conceptual transition is now taking place for dense neutrinos. Even the naming of \textit{collective flavor oscillations} of individual neutrinos in a self-consistent refractive field is increasingly being superseded by \textit{collective flavor conversions}. Indeed, these phenomena maintain very little of the historical oscillating character. The flavor oscillations, intended as the collective flavor waves in the plasma, are rarely excited individually, and rather form in most known cases an ensemble of waves with chaotic features which, in all other contexts, is usually characterized as \textit{turbulence}, described in terms of the dynamics of the collective field. The neutrino plasma is therefore not merely a gas of oscillating particles, but a coupled system of particles and emergent quasiparticles. In a way, this transition mirrors a rather common one in nuclear physics, to describe the physics in terms of the collective, rather than kinetic or single-particle, variables, ironically encoded in Weinberg's Third Law of Progress in Theoretical Physics~\cite{weinberg1983renormalization}: ``you may use any degrees of freedom you like to describe a physical system, but if
you use the wrong ones, you’ll be sorry.''

Recognizing the relevant degrees of freedom, however, is only the beginning. From the review given above it is clear that open questions have only multiplied, nested in decreasing level of abstraction and depth:

\begin{itemize}
    \item \textit{What are the limits of validity of the mean-field approximation underlying the quantum kinetic equations?} Numerical approaches are most likely not scalable to realistic setups, due to unavoidable limitations in the number of momentum modes available to neutrino occupation and in the number of neutrinos which can be simulated, either on a classical or a quantum computer. A theoretical path to concretely determine the dimensionless parameters deciding whether the mean-field assumption can be maintained would therefore be the way to address this question;
    
    \item \textit{What is the nature of instability in a generic plasma?} Beyond merely solving the dispersion relation, we should attempt to develop a qualitative feeling for the instability mechanism, so that we may directly predict when and why an instability arises. So far, a unified picture applies to most known cases: flavomon emission by flipped neutrinos with subdominant DLN. Fast, slow, and collisional instabilities then arise from different ways to satisfy the kinematics. We should remain alert to new mechanisms and forms of instability that may have been overlooked;
    
    \item \textit{How should the problem of flavor instability be formulated?} An initial-value problem for the coupled neutrino--matter--flavor dynamics is ideal, but generally intractable. Fast and slow instabilities have usually been treated, respectively, as local initial-value and boundary problems.  Collisional instabilities further blur this distinction, since collisions both shape the background and participate in the instability. What needs to be developed is a unified causal treatment amenable to theoretical understanding. The correct formulation may be connected with the absolute or convective nature of the instability, namely whether it grows at every point of space or only while the perturbation is convected away, in the latter case suggesting a boundary formulation;

    \item \textit{What is the nature of seeding?} Mass-induced mixing triggers flavor instabilities, but the amount of power that goes into inhomogeneous modes at the scale of the flavomon wavelength depends on the classical, hydrodynamical turbulence at these scales, well below the neutrino mean free path. It may also be that the increase in wavenumber caused by matter inhomogeneity in Eq.~\ref{eq:WKB_equations} allows large-scale fluctuations to develop into small-scale unstable modes. Thermal fluctuations may of course also provide a seeding mechanism. Finally, seeding by quantum fluctuations, i.e. from higher-order correlators in the BBGKY hierarchy is an additional, often forgotten contribution. This effect is not included in the standard kinetic equations, but present in the neutrino-flavomon kinetic equations, corresponding to the spontaneous flavomon emission. Which contribution is the most important remains an open question;
    
    \item \textit{What is the controlled kinetic description beyond the linear regime?} 
    The quasi-linear approximation has met some success, and has especially provided an intuitive understanding of saturation. However, its quantitative validity needs to be tested systematically. When may the waves be treated as weakly interacting quasiparticles? When is their evolution slow compared with their oscillation period? How should broad, overlapping, or nearly degenerate modes be treated? In an inhomogeneous medium, this description must moreover be joined to a WKB theory of flavor-wave propagation. Most importantly, if some of these assumptions break down, this should be understood not merely at the empirical level of numerical simulations. We should rather aim at a clear physical understanding of the reasons for this breakdown;
    
    \item \textit{What is the role of nonlinear wave--wave interactions?} Quasi-linear evolution retains the exchange of flavor between neutrinos and waves.  Nonlinear flavor-wave interactions may transfer excitation between scales, generate coherent structures, or produce a turbulent spectrum before wave--particle relaxation has removed the instability. It remains to be understood whether a quantitative theory of flavor-wave turbulence can be developed, what quantities such turbulence transports, and whether it exhibits cascades, universal spectra, or statistically stationary states. Just as importantly, genuine turbulence must be distinguished from fine-scale structure produced by phase mixing or kinematical dephasing alone;
    
    \item \textit{Why do many numerical approximations seemingly work so well?} Several numerical approximations, such as periodic boxes, global evolution with attenuation, and even in some cases global unresolved simulations, have shown remarkable numerical convergence based on empirical tests, yet a clear reason is lacking. We should therefore not only test convergence on selected examples, but understand why an approximation works, to determine in advance where it may fail. Beyond being encouraging, the success of these schemes should really be taken as a remarkable open question.
\end{itemize}

Of course, obtaining a reliable recipe for these effects is the most urgent practical goal of the field. Yet practicality itself has several levels. The final result must indeed be implementable in astrophysical simulations; but a fitted recipe is not, by itself, a physical understanding. A prescription reproducing the calculations from which it was inferred may fail when transported to a different environment. The decisive test is thus not postdiction, but prediction.

Numerical simulations will remain indispensable. For nonlinear systems, they play a role analogous to that of experiments: they reveal the outcome of the dynamics under controlled conditions. The remarkable progress of the last decade is therefore extremely promising. 

Following the scientific method, however, this empirical exploration should proceed hand in hand with interpretation: we must seek an intelligible framework to anticipate its outcomes.
This, too, is a practical endeavor. Theory should not develop in isolation, but should be guided by concrete phenomena that it must explain and, ultimately, predict. At the same time, we should resist constructing theories whose assumptions are tailored to the numerical experiments they are meant to explain; once again, we must predict, rather than postdict.

While the complexity of the nonlinear problem may be discouraging, one should not surrender the possibility of developing a clear intuition for it. In this sense, we may compare the situation of neutrino plasma kinetics to ordinary fluid mechanics, and its implementation in supernova simulations. A global supernova simulation is a masterpiece of numerical computation, yet we do not merely stare at its output: we recognize pressure-driven flows, buoyancy, convection, shocks, and turbulence, and can often anticipate their qualitative effects before the calculation is performed. That confidence rests on centuries of accumulated physical understanding. For the neutrino plasma---a genuinely unfamiliar state of matter---we are only beginning to identify the corresponding elementary processes.

The field will have reached a new level of maturity when collective flavor evolution becomes qualitatively predictable: when one can look at a neutrino plasma and understand what it is inclined to do before solving its equations in full. The numerical solution will still be needed to determine precisely how much conversion occurs, but it should be expressed in terms of the physical degrees of freedom evolving macroscopically, and especially it should no longer be needed to discover anew, in every example, \textit{why} conversion occurs. The central challenge for the coming years is thus not only to solve the equations of the neutrino plasma, but to learn how to read them.

\section*{Acknowledgments}
\addcontentsline{toc}{section}{\protect\numberline{}Acknowledgments}

I thank Huaiyu Duan, Julien Froustey, Luke Johns, Ian Padilla-Gay, Georg Raffelt, Sherwood Richers, Guenter Sigl, Irene Tamborra, Meng-Ru Wu, Zewei Xiong for several helpful and encouraging comments on this manuscript, and for many useful discussions. I also acknowledge support from the
Italian Ministero dell’Università e della Ricerca through the FIS 3 project FIS-2024-03087 (DD n. 18010 12-11-2025,
CUP E53C25002700001), and from the TAsP (Theoretical Astroparticle Physics) project.

\clearpage
\raggedbottom

\section*{Glossary}
\addcontentsline{toc}{section}{\protect\numberline{}Glossary}

{\footnotesize
\setlength{\abovedisplayskip}{0.4\baselineskip}
\setlength{\belowdisplayskip}{0.4\baselineskip}
\setlength{\abovedisplayshortskip}{0.2\baselineskip}
\setlength{\belowdisplayshortskip}{0.3\baselineskip}

\glossarymajorheading{Acronyms}

\begin{tabular}{@{}ll}
BBGKY & Bogoliubov--Born--Green--Kirkwood--Yvon hierarchy \\
BCS & Bardeen--Cooper--Schrieffer \\
CFC & Collective flavor conversion \\
CFI & Collisional flavor instability \\
DLN & Difference in lepton number \\
ECT & Effective classical transport \\
MSW & Mikheyev--Smirnov--Wolfenstein \\
NSM & Neutron-star merger \\
QKE & Quantum kinetic equation \\
QLT & Quasi-linear theory \\
SM & Standard Model \\
SN & Supernova (plural SNe) \\
WKB & Wentzel--Kramers--Brillouin \\
\end{tabular}

\vspace{0.7em}
\glossarymajorheading{Conventions}

\begin{tabular}{@{}L{0.20\linewidth}R{0.72\linewidth}@{}}
$\hbar=c=1$ &
Natural units are used throughout. \\[2pt]

$g^{\mu\nu}$ &
Metric convention $g^{\mu\nu}=\operatorname{diag}(+,-,-,-)$, so 
$K\cdot X=\Omega t-\bK\cdot\br$. \\[2pt]

$\displaystyle\sum_{\bp}$ &
Phase-space integral
$\displaystyle\int d^3\bp/(2\pi)^3$. \\[4pt]
$\displaystyle\sum_{v,w_E}$ &
Integral in axially symmetrical variables
$\displaystyle\int_{-1}^{+1} dv \int_{-\infty}^{+\infty} dw_E$, with $v=\bv\cdot \bn$ along an axis of symmetry $\bn$. Distributions in these variables are related to the canonical ones by
\[
D_{v,w_E}=\frac{E_\bp^2}{4\pi^2}\frac{dE_\bp}{dw_E}D_\bp=\frac{\delta m^6}{32\pi^2w_E^4}D_\bp.
\]
\\[4pt]

$\alpha,\beta$ &
Flavor indices. Unless stated otherwise, the discussion is restricted
to the two flavors $e$ and $\mu$. \\

$\rho_{\bp,\alpha\beta}(\br)$ &
Neutrino Wigner density matrix,
\[
\rho_{\bp,\alpha\beta}(\br)
 =\sum_\bq\,
 e^{i\bq\cdot\br}
 \left\langle
 a^\dagger_{\beta,\bp-\bq/2}
 a_{\alpha,\bp+\bq/2}
 \right\rangle .
\] \\[-2pt]

$\bar\rho_{\bp,\alpha\beta}(\br)$ &
Antineutrino Wigner density matrix,
\[
\bar\rho_{\bp,\alpha\beta}(\br)
 =\sum_\bq\,
 e^{i\bq\cdot\br}
 \left\langle
 b^\dagger_{\alpha,\bp-\bq/2}
 b_{\beta,\bp+\bq/2}
 \right\rangle .
\] \\[-2pt]
$\vec{P}_\bp$ & Polarization vectors; bold symbols are reserved for spatial vectors, arrows are used for flavor-space vectors.\\
$+i0$ & Retarded prescription for integration: contours should be deformed to pass below the resonant pole.\\

\end{tabular}

\vspace{1em}

\glossarymajorheading{Symbols}

\glossaryheading{Neutrinos and their flavor distribution}

\begin{tabular}{@{}L{0.20\linewidth}R{0.72\linewidth}@{}}

$P^\mu=(E_{\bp},\bp)$ &
Neutrino four-momentum, with $E_{\bp}=|\bp|$ in the
ultrarelativistic limit; the four-component velocity is $v^\mu=P^\mu/E_\bp$, although it does not transform as a four-vector. \\[2pt]

$\theta$, $v=\cos\theta$ & Polar angle and directional cosine relative to the specified symmetry or propagation axis.\\[2pt]

$\rho_{\bp},\bar\rho_{\bp}$ &
Neutrino and antineutrino flavor density matrices. For two flavors,
\[
\rho_{\bp}
 =\frac12
 \left(n_{\bp}
       +\vec{ P}_{\bp}\cdot\vec{\sigma}\right).
\]
$n_{\bp}$ is the total occupation, $P^z_{\bp}=D_\bp$ is
the DLN distribution, and
$\psi_{\bp}=P^x_{\bp}+iP^y_{\bp}$ is the flavor coherence; $\vec{\sigma}$ are the Pauli matrices. \\[4pt]

$\rho^\mu$, $\overline{\rho}^\mu$, $n^\mu$, $\overline{n}^\mu$ &
Momentum-integrated currents, e.g.\ $\rho^\mu=\sum_\bp \rho_\bp v^\mu$.\\[2pt]

$D^\mu$, $\psi^\mu$ &
Momentum-integrated flavor-isospin currents, e.g.\ $D^\mu=\sum_\bp (D_\bp-\overline{D}_\bp) v^\mu$; antineutrinos are included
with the opposite sign according to the flavor-isospin convention.\\[2pt]

$\langle\rho_{\bp}\rangle,
 \delta\rho_{\bp}$ &
Coarse-grained background and fluctuating component, with
$\rho_{\bp}=\langle\rho_{\bp}\rangle+\delta\rho_{\bp}$. \\
\end{tabular}

\glossaryheading{Neutrino dispersion and refractive potentials}

\begin{tabular}{@{}L{0.20\linewidth}R{0.72\linewidth}@{}}
${\sf\Omega}_{\bp}$ &
Matrix-valued effective energy, or flavor Hamiltonian, of a neutrino
with momentum $\bp$:
\[
{\sf\Omega}_{\bp}
 =E_\bp+{\sf\Omega}_{\bp,\rm vac}+{\sf\Omega}_{\bp,\rm mat}+{\sf\Omega_{\bp,\nu\nu}}.
\]
Flavor- and momentum-independent terms are omitted. \\[4pt]

$G_F$ &
Fermi constant. \\[2pt]

$\mu$ &
Characteristic neutrino--neutrino interaction strength,
$\mu=\sqrt{2}G_Fn_\nu$. \\[2pt]

$w_E$ &
Vacuum oscillation frequency,
$w_E=\delta m^2/(2E)$;$\delta m^2$ usually refers to the largest mass splitting. Antineutrinos are assigned negative $w_E$
in the flavor-isospin convention. \\[2pt]

$\theta_V,c_V,s_V$ &
Vacuum mixing angle and abbreviations
$c_V=\cos2\theta_V$ and $s_V=\sin2\theta_V$. \\[2pt]

$\lambda$ &
Charged-current matter potential,
$\lambda=\sqrt{2}G_Fn_e$. \\[2pt]

$u^\mu=(1,\bu)$ &
Four-component velocity of the matter background. \\[2pt]

$\Lambda^\mu$ &
Total flavor-dependent refractive shift, $\Lambda^\mu\equiv\lambda u^\mu+\sqrt{2}G_FD^\mu$.\\[2pt]

$\Delta E_{\bp}$, $\varepsilon_\bp$ &
Flavor-energy splitting of an individual neutrino mode,
$\Delta E_{\bp}
 =\Lambda\cdot v-w_Ec_V$, and total neutrino energy $\varepsilon_\bp=E_\bp + s \Delta E_\bp/2$ excluding flavor- and momentum-independent terms. Here $s=+1$ for $\nu_e$ and $\bar\nu_\mu$, and $s=-1$ for
$\nu_\mu$ and $\bar\nu_e$.\\
\end{tabular}

\glossaryheading{Flavor-wave frequency and wavevector}

\begin{tabular}{@{}L{0.20\linewidth}R{0.72\linewidth}@{}}
$X^\mu=(t,\br)$ &
Space-time coordinate. \\[2pt]

$K^\mu=(\Omega,\bK)$ &
Physical flavor-wave four-momentum, defined through
the Fourier dependence
$e^{-iK\cdot X}=e^{-i\Omega t+i\bK\cdot\br}$.\\[4pt]

$k^\mu=(\omega,\bk)$ &
Medium-shifted flavor-wave four-momentum,
$k^\mu\equiv K^\mu+\Lambda^\mu$. \\[4pt]

$\Omega_{\bK}$, $\omega_\bk$ &
Physical and shifted on-shell frequency of a flavomon, generally complex. \\[2pt]

$\gamma_{\bK}$ &
Flavomon growth or damping rate $\gamma_\bK=\mathrm{Im}\Omega_\bK$. Field amplitudes grow as
$e^{\gamma_{\bK}t}$, occupation numbers grow as
$e^{2\gamma_{\bK}t}$. \\[4pt]

$\Delta\varepsilon$ & Neutrino virtuality in flavomon emission/absorption
$\Delta\varepsilon
 =k\cdot v-w_Ec_V$.
On-shell emission requires $\Delta\varepsilon=0$. \\
\end{tabular}

\glossaryheading{Linear response and flavomons}
\begin{tabular}{@{}L{0.20\linewidth}R{0.72\linewidth}@{}}

$\psi^\mu$, $\psi^\mu_{\rm ext}$, $\tilde{\psi}^\mu=\psi^\mu+\psi^\mu_{\rm ext}$ & Flavor field sourced by neutrinos, external flavor field, and total flavor field.\\[2pt]
$\chi^\mu_\nu$ &
Flavor susceptibility, defined by the induced response $\psi^\mu=\chi^\mu_\nu\widetilde{\psi}^{\nu}$. \\[4pt]

$\varepsilon^\mu_\nu$ &
Flavor dielectric tensor, $\varepsilon^\mu_\nu
 =\delta^\mu_\nu-\chi^\mu_\nu$. \\[4pt]

 $\mathcal{S}^\mu$ & External seeding for flavomon field. \\[4pt]

$e^\mu$ &
Polarization eigenvector of a flavor-wave mode, satisfying
$\varepsilon^\mu_\nu e^\nu=0$
on shell. \\[4pt]

$\mathcal D^{\mu\nu}$ &
In-medium flavomon propagator,
\[
\mathcal D^{\mu\nu}
 =\frac{(\varepsilon^{-1})^{\mu\nu}}{\sqrt{2}G_F}.
\] \\[4pt]

$\Sigma^{\mu\nu}$ &
Flavomon self-energy, $
\Sigma^{\mu\nu}=\sqrt{2}G_F\chi^{\mu\nu}$. \\[4pt]

$\mathcal{Z}$ &
Wavefunction renormalization, or residue of the flavomon propagator,
\[
\mathcal{Z}^{-1}
 =e_\mu
  \left(\frac{\partial\varepsilon^\mu_\nu}{\partial\Omega}\right)
  e^\nu.
\] \\[4pt]

$N_{\bK}$ &
Occupation number of flavomons with wavevector $\bK$. \\[2pt]

$w(\bK,\bp)$ &
Transition kernel for the emission or absorption of a flavomon
with wavevector $\bK$ by a neutrino with momentum $\bp$. \\[2pt]

$\Gamma_{\nu_\alpha}(E)$ &
Incoherent interaction rate, or collisional spectral width, of a
neutrino of flavor $\alpha$ and energy $E$. \\[2pt]

$\epsilon$ &
Characteristic dimensionless DLN asymmetry,
schematically
\[
\epsilon\sim
\frac{n_{\nu_e}-n_{\nu_\mu}
      -n_{\bar\nu_e}+n_{\bar\nu_\mu}}{n_\nu}.
\]
The scale $\mu\epsilon$ sets the characteristic frequency of fast
flavor waves. \\
\end{tabular}

}

\addcontentsline{toc}{section}{\protect\numberline{}References}
\bibliographystyle{MyJHEP}
\bibliography{references}

\end{document}